\documentclass[twocolumn, amsmath,amssymb,aps, prx,
showkeys,showpacs, floatfix, allowpagebreak] {revtex4-2}
\usepackage{graphicx}
\usepackage[usenames,dvipsnames]{color}
\usepackage{bm}
\usepackage[colorlinks=true,breaklinks=true,allcolors=blue]{hyperref}

\def\re{{\rm e}}
\def\ri{{\rm i}}

\begin{document}

\title{Quantum Electronic Circuits for Multicritical Ising Models}
\author{Ananda Roy}
\email{ananda.roy@physics.rutgers.edu}
\affiliation{Department of Physics and Astronomy, Rutgers University, Piscataway, NJ 08854-8019 USA}

\begin{abstract}
Multicritical Ising models and their perturbations are paradigmatic models of statistical mechanics. In two space-time dimensions, these models provide a fertile testbed for investigation of numerous non-perturbative problems in strongly-interacting quantum field theories. In this work, analog superconducting quantum electronic circuit simulators are described for the realization of these multicritical Ising models. The latter arise as perturbations of the quantum sine-Gordon model with $p$-fold degenerate minima, $p =2, 3,4,\ldots$. The corresponding quantum circuits are constructed with Josephson junctions with $\cos(n\phi + \delta_n)$ potential with $1\leq n\leq p$ and $\delta_n\in[-\pi,\pi]$. The simplest case, $p = 2$, corresponds to the quantum Ising model and can be realized using conventional Josephson junctions and the so-called $0-\pi$ qubits. The lattice models for the Ising and tricritical Ising models are analyzed numerically using the density matrix renormalization group technique. Evidence for the multicritical phenomena are obtained from computation of entanglement entropy of a subsystem and correlation functions of relevant lattice operators. The proposed quantum circuits provide a systematic approach for controlled numerical and experimental investigation of a wide-range of non-perturbative phenomena occurring in low-dimensional quantum field theories. 
\end{abstract}

\maketitle

\section{Introduction}
\label{sec:intro}
Quantum simulation~\cite{Feynman_1982, Lloyd1997} is an indispensable technique for investigation of strongly-interacting quantum field theories~(QFTs)~\cite{Jordan2011, Jordan2012}. With the advent of noisy intermediate-scale quantum simulators and algorithms, gate-based digital quantum simulation has been used to investigate lattice models for a wide-range of non-perturbative QFT problems. These include simulation of quantum many-body dynamics~\cite{Smith2019}, topological phase-transitions~\cite{Smith2022} and confinement in perturbed Ising models~\cite{Vovrosh2021, Tan_2021, Lamb2023}. However, given the number of available qubits and the coherence properties of existing quantum simulators, generalization of the aforementioned simulation protocols to investigate generic QFTs with thousands, potentially millions, of qubits remains a daunting challenge in the near term. Such large system sizes are necessary for most QFT problems since the convergence of the lattice model to the scaling limit is usually slow, examples being QFTs describing quantum critical points and their vicinities. 

\begin{figure}
    \centering
    \includegraphics[width=0.5\textwidth]{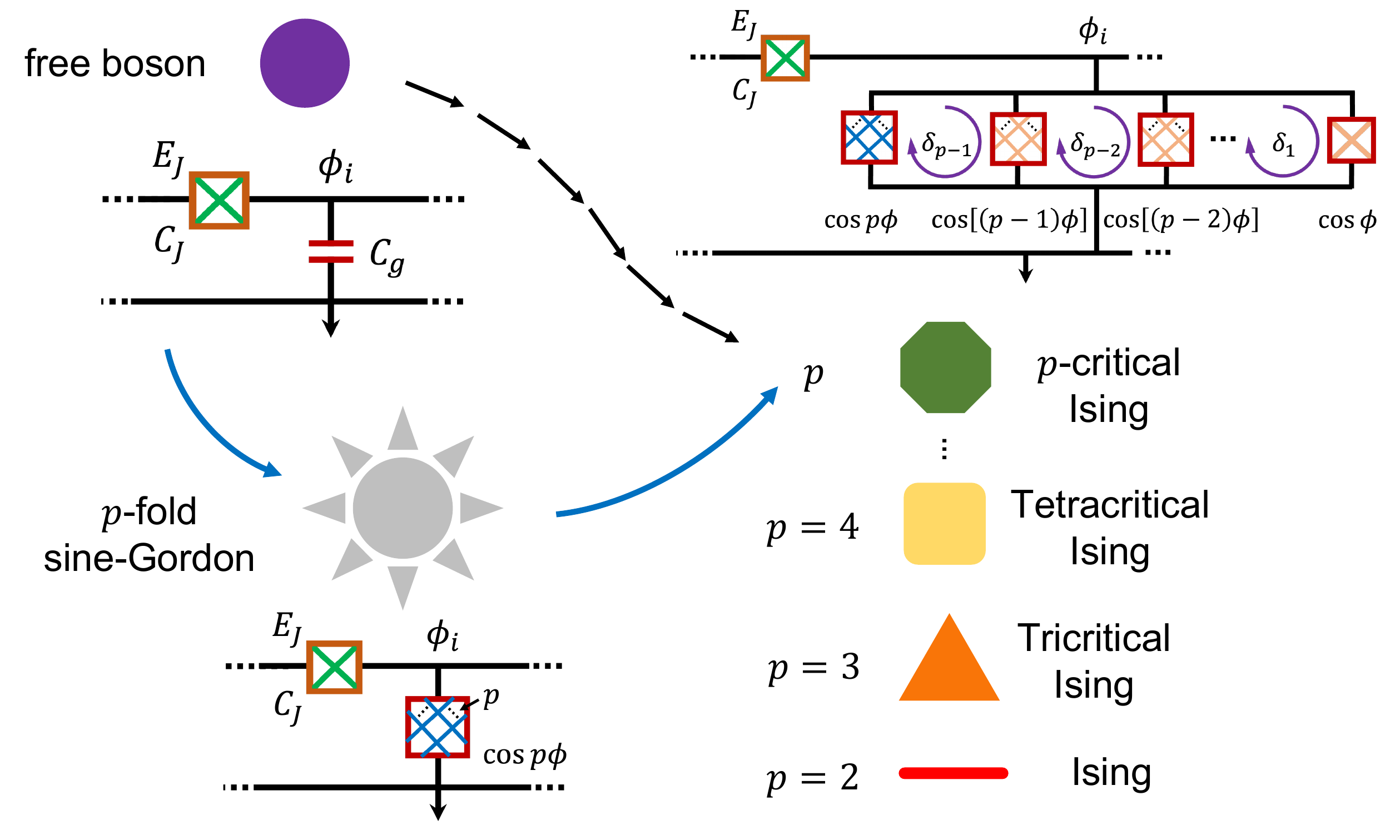}
    \caption{ \label{fig:schematic} Quantum circuit scheme for realization of multicritical Ising models. The latter occur as infrared fixed points of the renormalization group flow trajectory~(black arrows) starting from the free compactified boson QFT in the ultraviolet. The blue arrows indicate the logic of the circuit construction. The corresponding unit cells of the quantum circuit are shown. In all cases, the horizontal link contains a Josephson junction~(junction energy $E_J$ and junction capacitance $C_J$). The node flux at the $i^{\rm th}$ site is indicated. The circuit element on the vertical link determines the nature of the QFT. The latter element is a capacitor~[$\cos(p\phi)$ Josephson junction] for the free boson model~[sine-Gordon model with $p$ degenerate minima]. A parallel circuit of $\cos(n\phi)$ Josephson junctions with $n = 1,2, \ldots, p$ on the vertical link realizes the $p-$critical Ising model. The~$p-1$ phase-differences between the different circuit elements, denoted by $\delta_n$, $n = 1,\ldots, p-1$, are selected depending on the specific model.} 
\end{figure}

Analog quantum simulation~\cite{Doucot2004, Cirac2010, Buchler2005, Casanova2011, Roy2019} provides a near-term, more tractable alternative to the aforementioned digital approach. This is particularly relevant for the investigation of those QFT problems which require probing properties at longer length-scales than that permitted using current digital quantum simulators. Indeed, analog simulation has had considerable success in probing complex quantum many-body systems with simulators based on trapped atoms~\cite{Duca2015, Parsons2016, Gross2017, Bernien2017, Samajdar2020, Ebadi_2021}, trapped ions~\cite{Friedenauer2008, Kim2010} and superconducting quantum electronic circuits~(QECs)~\cite{Kuzmin2018, Kuzmin2019, Leger2019, PuertasMartinez2019}. In this work, we focus on QEC-based quantum simulators in two space-time dimensions. These QEC simulators rely on the robust, tunable, dispersive Josephson nonlinearity to give rise to strongly-interacting nonlinear QFTs~\cite{Roy2020b, Roy2023}. In fact, investigation of arrays with thousands of {\it quantum} Josephson junction have already been experimentally performed~\cite{Kuzmin2019, Leger2019}.

In a QEC array, the fundamental lattice degree of freedom is the superconducting phase at a lattice site. This provides a convenient starting point for discretization of a wide range of bosonic QFTs realizable as perturbations of the free boson QFT. The latter occurs naturally as a long-wavelength description of a one-dimensional Josephson junction array in the limit of large Cooper-pair tunneling strength~\cite{Bradley1984, Glazman1997}. In the continuum limit, the bosonic field arises from the corresponding lattice superconducting phase after coarse-graining. This approach has been utilized to give rise to the quantum sine-Gordon~(sG) model~\cite{Roy2020b}, a non-integrable, two-frequency sG model~\cite{Roy2023} and a multi-field generalization of the sG model~\cite{Roy2019}. 

In contrast to the earlier proposals which considered perturbations of the free boson QFT that lead to flows to strongly-interacting massive QFTs, this work describes QEC simulators that realize quantum critical points of varying universality classes. In particular, we analyze QEC simulators which, in the scaling limit, are described by multicritical Ising models. The latter are diagonal, unitary, minimal models of conformal field theories~\cite{diFrancesco1997}. These have played a central role in the understanding of two dimensional, critical, classical statistical mechanics models~\cite{Belavin1984} and are the starting point for systematic analysis of perturbed, integrable or otherwise, conformal field theories~\cite{Zamolodchikov1987, Zamolodchikov1989}. Furthermore, stacks of such critical models, with appropriate couplings, give rise to a large class of topological phases~\cite{Teo2014, Mong2014} relevant for topological quantum computation~\cite{Kitaev2003}. 

It is well-known that these multicritical Ising QFTs arise in the scaling limit of restricted solid-on-solid~(RSOS) models~\cite{Andrews:1984af}. The corresponding quantum Hamiltonians, owing their integrable construction, have been analyzed extensively using Bethe ansatz~\cite{Bazhanov_1990}. The tricritical Ising model has been shown to occur also in the Blume-Capel model~\cite{Blume1966, Capel1966} and in interacting chains of Majorana zero modes~\cite{Rahmani2015, Brien2018}. However, unlike RSOS models, there is no clear path towards generalization of the Blume-Capel model or the Majorana chains to generic multicritical Ising models. The goal of this work is to present QEC lattices that have the same versatility as the RSOS models while having the merit of being potentially realized in an experiment. 

Starting with the QEC lattice model for the quantum sG field theory with $p$-fold degenerate minima, $p = 2, 3, \ldots$, perturbations of the form~$\cos(n\phi  + \delta_n)$ are systematically added. Here, $1\leq n<p$ and $\delta_n\in[-\pi,\pi]$. As shown in this work, these lattice models give rise to multicritical Ising models upon appropriate choice of parameters. In contrast to the RSOS models, the proposed lattice models are non-integrable, even though they give rise to the integrable multicritical Ising QFTs in the scaling limit. Due to their non-integrable nature, these lattice models are analyzed numerically, using the density matrix renormalization group~(DMRG) technique~\cite{DMRG_TeNPy}. 

Note that similar non-integrable lattice models could be conceived starting with the XYZ spin chain regularization of the quantum sG model~\cite{Lukyanov2003, Baxter2013}. In comparison to the models proposed in this work which use only nearest-neighbor interactions, the generalizations of the XYZ chain would require longer-range interactions between the spins. Furthermore, the generalized XYZ models suffer from larger corrections to scaling compared to the proposed models~(see Ref.~\cite{Roy2020b} for a numerical demonstration). As a result, QEC circuits are a more suitable platform for realization of the perturbed sG models considered in this work. 

The article is organized as follows. Sec.~\ref{sec:gen_appr} describes the general scheme for realizing arbitrary multicritical Ising models. Secs.~\ref{sec:Ising} and~\ref{sec:TCI} are devoted to the Ising and the tricritical Ising models respectively. Sec.~\ref{sec:concl} provides a concluding summary and outlook. 

\section{General Scheme}
\label{sec:gen_appr}
The main idea behind the QEC realization of multicritical Ising models is based on the well-known notion that the diagonal, unitary minimal models of conformal field theories arise as multicritical points of an effective Ginzburg-Landau action~\cite{Zamolodchikov:1986db}. To arrive at this effective action, we consider perturbations of the euclidean quantum sG action with $p$-fold degenerate minima describing the scalar field $\varphi$:
\begin{align}
\label{eq:psG}
{\cal A} &= \int d^2x\left[\frac{1}{16\pi}(\partial_\nu\varphi)^2 - 2\mu\cos(\beta\varphi)\right]\nonumber\\&\quad - \sum_{n = 1}^{p-1}2\lambda_n\int d^2x \cos\left(\frac{n\beta\varphi}{p} + \delta_{n}\right),
\end{align}
where $\mu, \lambda_n$-s are coupling constants and $\delta_n$-s are suitably chosen phases. The case $\lambda_n = 0\ \forall\ n$ corresponds to the ordinary sG model with $p$-fold degenerate minima with coupling constant $\beta$, where we consider the case~$0\leq\beta^2\leq1$. Appropriate choices of $\{\lambda_n,\delta_n\}$ induces a flow to the quantum critical points of multicritical Ising universality class. The latter are characterized by the central charges
\begin{equation}
\label{eq:c}
c_p = 1-\frac{6}{(p+1)(p+2)},\ p = 2,3,\ldots.
\end{equation}

The perturbed sG action of Eq.~\eqref{eq:psG} describes the scaling limit of the QEC lattice shown in Fig.~\ref{fig:schematic}~(top right).  Each unit cell has a Josephson junction on the horizontal link with junction energy,~$E_J$, and junction capacitance,~$C_J$. The vertical link contains the most crucial circuit element. When the latter is an ordinary capacitor with capacitance $C_g$, the QEC array can be in a superconducting phase~\cite{Bradley1984, Glazman1997}. This happens when $E_J>E_c$, where $E_c = (2e)^2/2C_g$~\cite{Roy2020a}. The long-wavelength properties of the array are described by the free, compactified boson QFT. Choosing the circuit element on the vertical link to be a $\cos(p\phi)$ Josephson junction gives rise to the quantum sG model with $p$-fold degenerate minima~(see Ref.~\cite{Roy2023} for the analysis of~$p = 2$). Finally, the parallel circuit configuration shown in Fig.~\ref{fig:schematic}~(top right) gives rise to the perturbed sG model of Eq.~\eqref{eq:psG} in the scaling limit. 

The Hamiltonian of the QEC array with $L$ sites and periodic boundary conditions is given by
\begin{align}
\label{eq:H_circ}
H &= E_c\sum_{j = 1}^Ln_j^2 + \epsilon E_c\sum_{j = 1}^{L}n_jn_{j+1}\nonumber\\&\quad  - E_g\sum_{j = 1}^Ln_j- E_J\sum_{j =1}^L\cos(\phi_j - \phi_{j+1})
\nonumber\\&\quad- \sum_{n = 1}^{p}\sum_{j = 1}^LE_{J_n}\cos(n\phi_j + \delta_n),
\end{align}
where $\delta_p$ is set to 0. Here, $n_j$, the excess number of Cooper-pairs on each island, and $\phi_k$, the node-flux, are canonically conjugate satisfying: $[n_j, \re^{\pm\ri\phi_k}] = \pm\delta_{jk}\re^{\pm\ri\phi_k}$. The nearest-neighbor interaction proportional to $\epsilon\leq1$ arises due to the capacitance $C_J$~\cite{Glazman1997}. In this work, we chose $\epsilon = 0.2$ and~$E_g/E_c = 1.2$~\cite{Roy2020b, Roy2023}, but similar results could be obtained for other choices.  Furthermore, the numerical computations were performed by choosing $E_c = 1.0$. The third and fourth terms arises due to a gate voltage at each node taken to be uniform and the coherent tunneling of Cooper-pairs. The last term in Eq.~\eqref{eq:H_circ} arises from the parallel circuit arrangement shown in Fig.~\ref{fig:schematic}(b). Setting $E_{J_n} = 0$ for $1\leq n\leq p$ and $1\leq n\leq p-1$ in the Eq.~\eqref{eq:H_circ}, in the scaling limit, gives rise to the free boson and the quantum sG model with $p$-fold degenerate minima respectively. In general, tuning the $E_{J_n}$-s and $\delta_n$-s leads to the QEC array being in a gapless state with the universality class of the critical point determined by $p$. The Ising and tricritical Ising cases are described below. 

\section{The Ising model}
\label{sec:Ising}
The simplest realization of the Ising model from a perturbed sG model is obtained by choosing $p = 2$ and $\delta_1 = \pi/2$. For the continuum model, this phase-transition has been analyzed using form-factors~\cite{Delfino1998}, semi-classical methods~\cite{Mussardo2004} and truncated conformal space approaches~\cite{Bajnok2000}. The existence of the Ising critical point can be straightforwardly inferred already from the classical potential. The latter is obtained by rescaling the fields $\tilde\varphi = \beta\varphi$ in the limit $\beta\rightarrow0$. The potential is $V(\tilde\varphi) = -2\mu\cos\tilde\varphi - 2\lambda_1\sin(\tilde\varphi/2)$. For $\lambda_1 < 4\mu$, the $V(\tilde\varphi)$ has two degenerate minima characteristic of the ferromagnetic phase of the Ising model. The minima occur at $\varphi_0$ and $2\pi - \varphi_0$, where $\varphi_0$ is some classical field minima. The latter two values are related by a  $\mathbb{Z}_2$ symmetry operator (see below for the construction of this operator for the QEC model). These two minima coalesce at $\lambda_1 = 4\mu$ indicating an Ising-type phase-transition. Further increase of $\lambda_1/\mu$ results in a non-degenerate potential minimum, as is expected from the paramagnetic phase of the Ising model. Obviously, the actual value of the ratio $\lambda_1/\mu$ when the phase-transition occurs is different for finite $\beta^2$~(see Fig.~\ref{fig:Ising_pd} for DMRG results).  

The corresponding circuit Hamiltonian to realize the Ising model is obtained by setting $p=2, \delta_1 = \pi/2$ in Eq.~\eqref{eq:H_circ}. The circuit element on the vertical link is a parallel circuit of a conventional Josephson junction~(junction energy~$E_{J_1}$) and a~$\cos2\phi$ Josephson junction~(also known as the~$0-\pi$ qubit~\cite{Doucot2002, Ioffe2002, Kitaev2006c, Brooks2013, Gladchenko2008, Smith2020, Gyenis2021}) with a magnetic flux threading the loop~\footnote{Note that the effective capacitance~$C_g$ is the sum of the two capacitances of the~$\cos\phi$ and the~$\cos2\phi$ Josephson junctions.}.  The different couplings are chosen as follows. First, $E_J/E_c$ and $E_g/E_c$ are chosen such that the QEC array is in the superconducting~(free boson) phase when $E_{J_n} = 0$ $\forall n$. Subsequently, the parameters $E_{J_1}, E_{J_2}$ are chosen to give rise to the perturbed sG model in the scaling limit. The sG coupling is given by:~$\beta^2  = K/2$, where $K$ is the Luttinger parameter of the free boson theory. The relationship between the lattice and the continuum parameters can be summarized as~(see Supplementary Materials of Ref.~\cite{Roy2023} for details):
\begin{align}
\mu =C\ E_{J_2} E_c^{1-2\beta^2},\  \lambda_1=C'\ E_{J_1} E_c^{1-\beta^2/2}, 
\end{align}
where $C, C'$ are non-universal functions of $\beta^2$, which can, in principle, be determined numerically. Note that the~$\mathbb{Z}_2$ symmetry for the potential term~[the last term of Eq.~\eqref{eq:H_circ}] is~$\phi_j\rightarrow\pi - \phi_j$,~$j = 1,\ldots, L$. The corresponding symmetry operator that performs the transformation~$\phi_j \rightarrow \pi-\phi_j$ is:
\begin{equation}
\label{eq:O_def}
{\cal O} = \left(\prod_{j = 1}^L \re^{\ri\pi n_j}\right){\cal C},
\end{equation}
where ${\cal C}$ is the charge-conjugation operator that acts as
\begin{equation}
\label{eq:C_act}
{\cal C}\re^{\pm \ri\phi_j}{\cal C} = \re^{\mp \ri\phi_j},\ {\cal C}n_j{\cal C} = -n_j. 
\end{equation}
Using Eq.~\eqref{eq:C_act}, it is straightforward to show that~${\cal O}$ is indeed a symmetry of the potential term. Notice however that it is not a symmetry of the lattice Hamiltonian unless $E_g = 0$. Nevertheless, as will be numerically demonstrated below, the symmetry associated with~${\cal O}$ emerges in the scaling limit also for $E_g\neq0$. In the scaling limit, the QEC lattice operators that correspond to the primary fields of the Ising model are:
\begin{equation}
\label{eq:Ising_ops}
\sigma\sim\cos\phi_j + \ldots,\ \varepsilon\sim \sin\phi_j + \ldots
\end{equation}
where the dots correspond to subleading corrections. The two fields have scaling dimensions $1/8$ and $1$ respectively. This is verified below in the numerical analysis. Furthermore, a Jordan-Wigner mapping from the Ising spin chain to the free-fermion Hamiltonian leads to identification of the lattice fermion operators as $(\prod_{k<j}\re^{\ri\pi n_k})\re^{\ri\phi_j}$. It is straightforward to check the fermionic nature of these nonlocal operators using the commutation relation of~$\phi_j$ and~$n_k$ given below Eq.~\eqref{eq:H_circ}. Notice that these nonlocal operators are the soliton creation operators of the parent sine-Gordon model with two-fold degenerate minima. These soliton-creation operators have Lorentz spin 1/2 and thus, also exhibit fermionic statistics~\cite{Lukyanov2001}.

\begin{figure}
    \centering
    \includegraphics[width=0.5\textwidth]{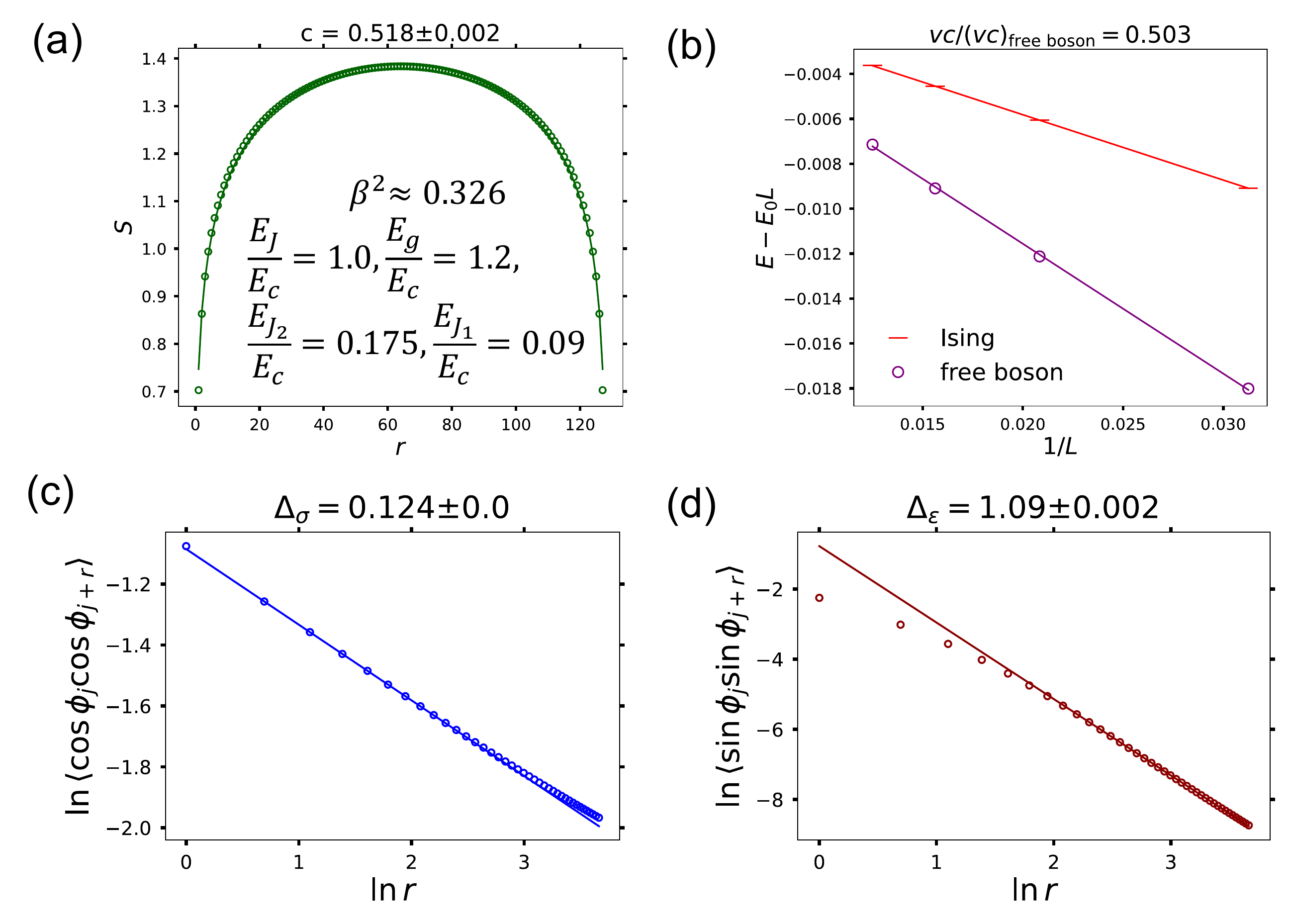}
    \caption{ \label{fig:Ising_1} DMRG results for the Ising transition. The different couplings were chosen as shown on the top left panel. (a) A QEC array of $L = 128$ sites with periodic boundary conditions was analyzed. The entanglement entropy,~$S$, as a function of subsystem size~$r$ exhibits a characteristic logarithmic dependence~[Eq.~\eqref{eq:S_r}]. The obtained central charge is close to the expected value of~$1/2$~[set~$p = 2$ in Eq.~\eqref{eq:c}]. The discrepancy with the expected value is due to finite size effect.~(b) The product of the central charge and the `Fermi' velocity is determined from the scaling of the Casimir energy~[Eq.~\eqref{eq:cas}]. The ratio of this product is close to 1/2 as expected.~(c, d) Infinite DMRG results for the algebraic decay of the lattice operators corresponding to the Ising field operators~$\sigma, \varepsilon$~[see Eq.~\eqref{eq:Ising_corr}]. The small discrepancy with the expected values of $(\Delta_\sigma, \Delta_\varepsilon) = (1/8, 1)$ is due to the difficulty of locating the exact critical point as well as finite truncation errors occurring in the infinite DMRG simulation.  } 
\end{figure}

To unambiguously obtain the critical point, entanglement entropy,~$S$, was computed using DMRG for a QEC array of $L = 128$ sites with periodic boundary conditions~(see Fig.~\ref{fig:Ising_1} for lattice parameters). At the critical point, standard results of conformal field theory predict~\cite{Calabrese2004}
\begin{equation}
\label{eq:S_r}
S(r) = \frac{c}{3}\ln\left[\frac{L}{\pi a}\sin\frac{\pi r}{L}\right] + S_0,
\end{equation}
where $a$ is the lattice spacing and $S_0$ is some non-universal constant. Here, $c$ is the central charge and equals 1/2~[set $p = 2$ in Eq.~\eqref{eq:c}]. The obtained value of $c$ is close to 1/2~[see Fig.~\ref{fig:Ising_1}(a)] with the discrepancy arising due to rather strong finite-size effects. This was verified by simulating different system sizes~$L = 40, 64,$ and 128. As an additional check of the numerical results, the scaling of the ground-state energy is computed with system size. For periodic boundary conditions, this obeys~\cite{diFrancesco1997}
\begin{equation}
\label{eq:cas}
E = E_0L -\frac{\pi cv}{6L} + {o}(1/L),
\end{equation}
where $v$ is the `Fermi' velocity. Since the additional sine and cosine perturbations do not renormalize~$v$, the ratio of $vc$ computed for the Ising and the free-boson critical points should be 1/2. This is verified in Fig.~\ref{fig:Ising_1}(b). Note that the simulating the QEC lattice Hamiltonian requires manipulating Hilbert space dimension of $17$ at each site~(see Supplementary Materials of Ref.~\cite{Roy2023} for details), which made analysis of larger system sizes rather challenging. The single-site DMRG implementation of the TeNPy package was used throughout this work.

At the critical point, the correlation functions of lattice operators~$\cos\phi_j$ and~$\sin\phi_j$ are computed using infinite DMRG. These correlation functions are algebraic:
\begin{equation}
\label{eq:Ising_corr}
\langle \cos\phi_j\cos\phi_{j+r}\rangle\propto \frac{1}{r^{2\Delta_\sigma}}, \ \langle \sin\phi_j\sin\phi_{j+r}\rangle\propto \frac{1}{r^{2\Delta_\varepsilon}}.
\end{equation}
The obtained results for the above correlation functions are shown in Fig.~\ref{fig:Ising_1}(c,d). The small discrepancy between the obtained values of $\Delta_\sigma, \Delta_\epsilon$ and the predicted ones are  due to the uncertainty in the location of the critical point as well as finite entanglement truncation~\cite{Pollmann2009}. 
\begin{figure}
    \centering
    \includegraphics[width=0.5\textwidth]{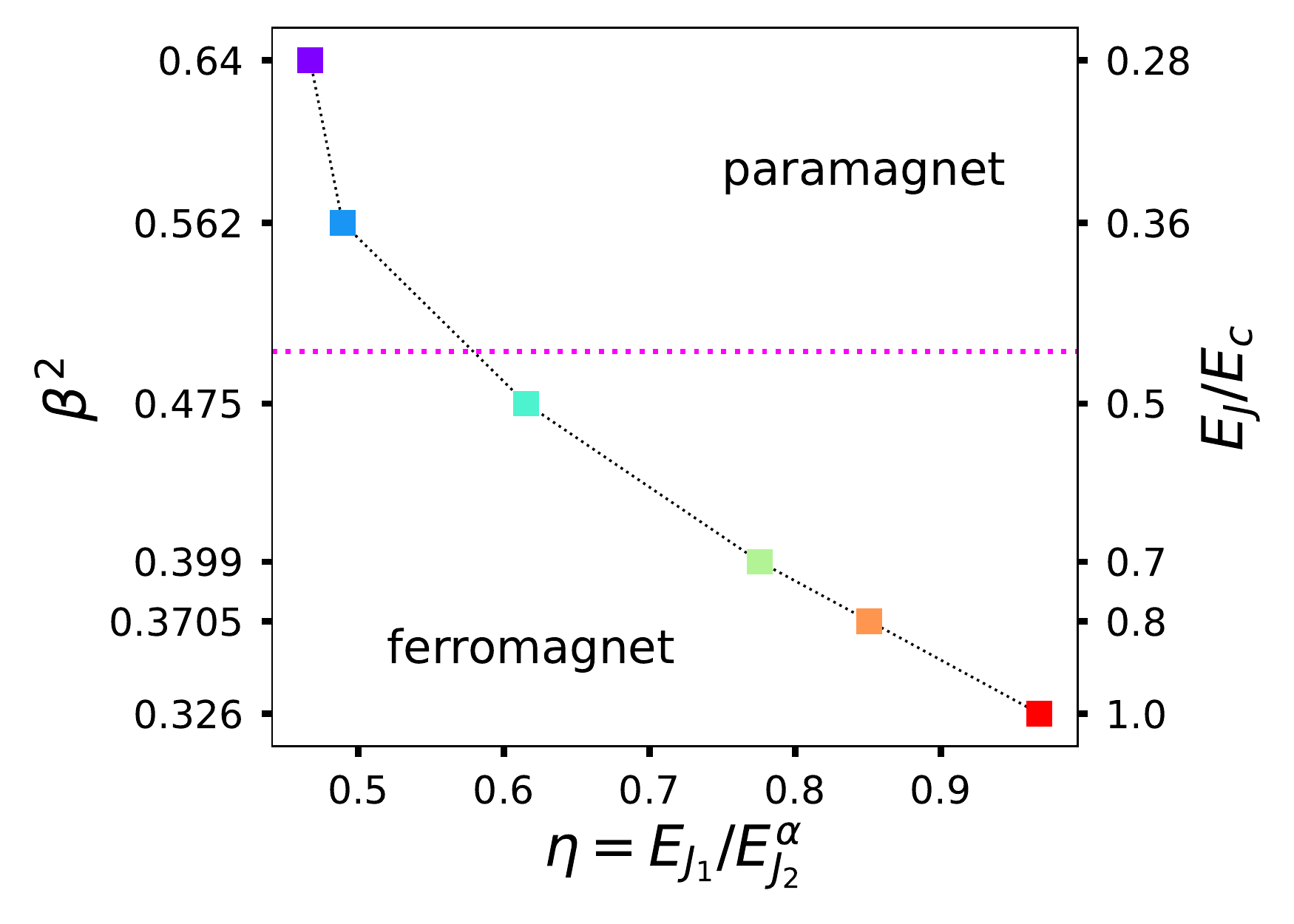}
    \caption{\label{fig:Ising_pd} DMRG results for the phase-diagram associated with the Ising transition as a function of the dimensionless coupling~$\eta = E_{J_1}/E_{J_2}^\alpha$ and~$\beta^2$. Here,~$\alpha = (1-\beta^2/4)/(1-\beta^2)$. The results were obtained for a periodic QEC array with $L = 128$ and $E_{J_2}/E_c = 0.175$. The solid markers correspond to the numerically obtained location of the phase-transition, while the black dashed line is a guide to the eye. The magenta dotted line corresponds to the free-fermion point~($\beta^2 = 1/2$) of the sG model. For a fixed~$\beta^2$, increasing~$\eta$ induces an Ising phase-transition from the ferromagnetic to the paramagnetic phase. The locations of the critical points are determined with a precision of $E_{J_1}/E_{J_2} = 0.001$~(the corresponding error bars are too small to be visible). } 
\end{figure}

Fig.~\ref{fig:Ising_pd} shows the phase-diagram around the Ising transition as a function of a dimensionless variable~$\eta = E_{J_1}/E_{J_2}^\alpha$ and~$\beta^2$. Here, $\alpha = (1-\beta^2/4)/(1-\beta^2)$. The values of $E_J/E_c$ corresponding to the different choices of~$\beta^2$ are shown on the right y-axis. For a given choice of~$\beta^2$, increasing $\eta$ induces the Ising phase-transition from the ferromagnetic to the paramagnetic phase~(see also Fig. 12 of Ref.~\cite{Bajnok2000} for a truncated conformal space analysis performed for $\beta^2<1/2$). The location of the critical point was obtained by sweeping the coupling ratio $E_{J_2}/E_{J_1}$ for different choices of $E_J/E_c$ for a periodic chain of $L= 128$ sites and computing the central charge as in Fig.~\ref{fig:Ising_1}. 

\begin{figure}
    \centering
    \includegraphics[width=0.5\textwidth]{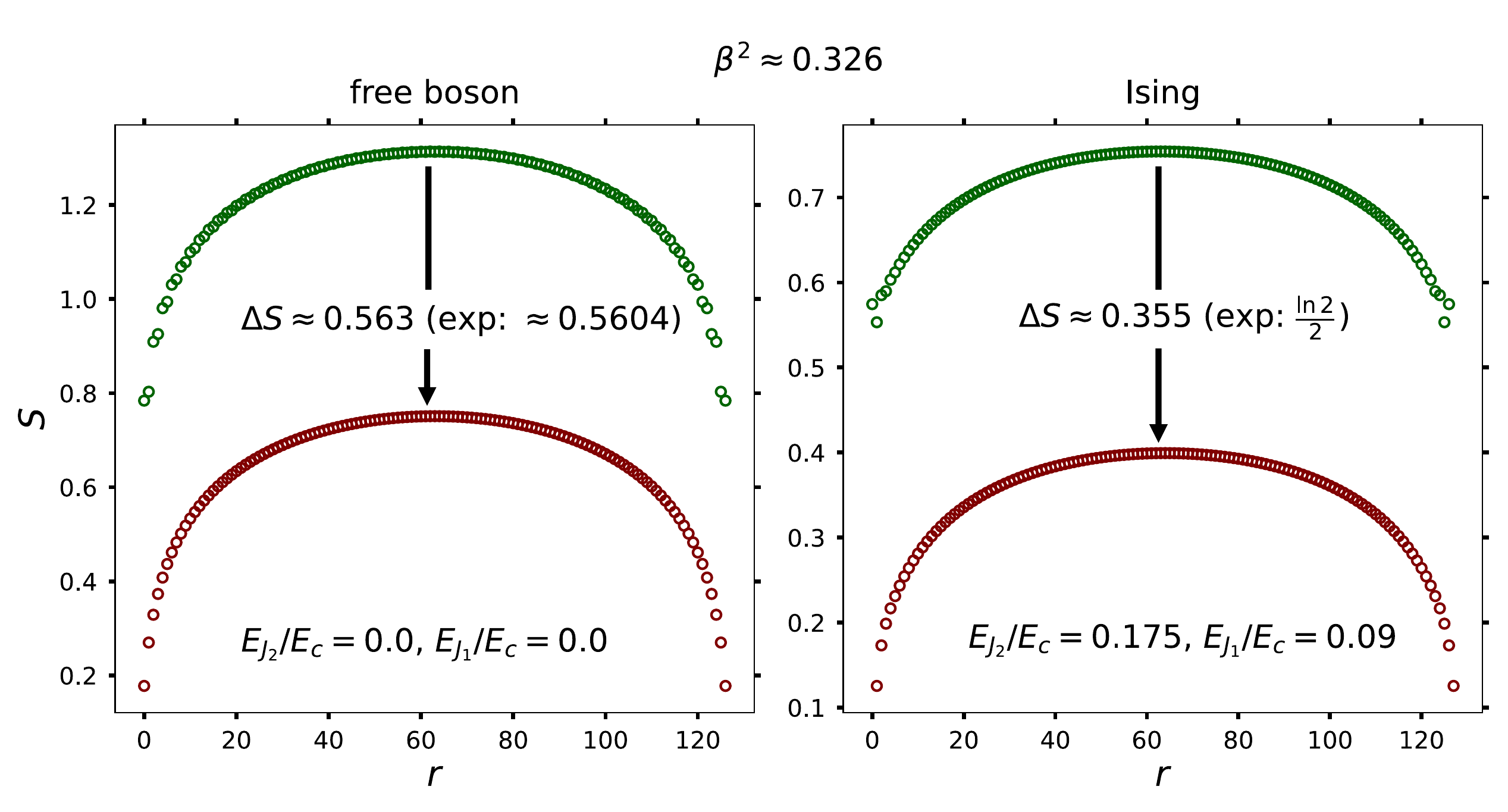}
    \caption{ \label{fig:Ising_flow} DMRG results for the boundary renormalization group flows for the free boson~(left) and the Ising~(right) models. The parameters, $E_J, E_g$ were chosen as in Fig.~\ref{fig:Ising_1}. In contrast to the rest of the paper, in this computation, the interaction terms of Eq.~\eqref{eq:H_circ} between sites $L$ and 1 are absent. Choosing~$E_{J_n} = 0$~$\forall n$ realizes the free boson model with free boundary condition with~$\beta^2\approx0.326$. The corresponding entanglement entropy is shown in dark green for different sizes of the subsystem. To realize the fixed boundary condition, we chose~$E_{J_b}/E_c = 1$  in the numerical simulations. The corresponding entanglement entropies are shown in maroon. The change in the boundary entropy in the infrared is obtained by taking the difference between the two curves at the center of the chain. The obtained and expected results are shown. Now, turning on the couplings~$E_{J_1}, E_{J_2}$ throughout the array realizes the Ising model~[see Fig.~\ref{fig:Ising_1}]. This Hamiltonian without~(with) the same boundary potential realizes the Ising model with free~(fixed) boundary condition. The corresponding entanglement entropies with the change associated with the change in the boundary condition are shown in the right panel. } 
\end{figure}

With the aforementioned phase-diagram determined, it is straightforward to consider the different perturbations of the Ising critical point. The `thermal' perturbation has been analyzed here and involves tuning the ratio $E_{J_1}/E_{J_2}$ away from the Ising critical point. The `magnetic' perturbation~\cite{Zamolodchikov1989} could be analyzed by adding a `longitudinal field' to the Hamiltonian of Eq.~\eqref{eq:H_circ}. This involves adding an extra $-\sum_{j = 1}^L\cos\phi_j$ term to the QEC Hamiltonian. As a simple application, consider the case of the free-boson and the Ising models with boundary fields~\cite{Ghoshal1994, Fendley1994}. First, in the QEC Hamiltonian~[Eq.~\eqref{eq:H_circ}], with~$E_{J_n} = 0$~$\forall n$, boundary fields~$-E_{J_b}\cos\phi_j$ are turned on at sites $j = 1, L$. In the continuum limit, this corresponds to the boundary sine-Gordon model, with a boundary potential $\cos\beta\varphi/2$ added to the euclidean action of Eq.~\eqref{eq:H_circ}. The boundary potential induces a flow from the free to the fixed boundary condition. The change in the boundary condition manifests itself in a change in the boundary entropy~\cite{Affleck1991}. The latter can be measured by computing the change in the subleading~$O(1)$ term in the entanglement entropy. The change for the free-boson case is well-known and given by $-(\ln\beta^2)/2$~(see, for example, Refs.~\cite{Saleur1998, Roy2020a}). The DMRG results are shown in Fig.~\ref{fig:Ising_flow}, left panel for~$\beta^2\approx0.326$. Next, upon turning on $E_{J_1}, E_{J_2}$ to the values which realize the critical Ising model in the bulk while keeping all other parameters the same, the QEC array realizes the Ising model with or without a longitudinal boundary field. The corresponding change in boundary entropy is~$(\ln2)/2$~\cite{Affleck1991}. The DMRG results are shown in Fig.~\ref{fig:Ising_flow}, right panel. 
 
\section{The Tricritical Ising model}
\label{sec:TCI}
The tricritical Ising model is the next in the series of models that can be realized with QECs. Consider the case $p = 3$ and $\delta_1 =0$, $\delta_2 = \pi$ in Eqs.~(\ref{eq:psG}, \ref{eq:H_circ}). For $E_{J_1} = E_{J_2} = 0$, the lattice model of Eq.~\eqref{eq:H_circ} realizes the sine-Gordon model with three degenerate minima. Similar analysis as in the Ising case leads to $\beta^2 = 9K/8$, where $K$ is the Luttinger parameter of the parent free boson theory. First, consider the classical potential:
\begin{equation}
V(\tilde\varphi) = -2\mu\cos\tilde\varphi - 2\lambda_1\cos\frac{\tilde\varphi}{3} + 2\lambda_2\cos\frac{2\tilde\varphi}{3}.
\end{equation}
Straightforward computation yields a critical Ising line 
\begin{equation}
\lambda_2 = \frac{\lambda_1}{4} + \frac{9\mu}{4}
\end{equation}
terminating at a tricritical Ising point for $\lambda_1 = 15\mu, \lambda_2 = 6\mu$. The phase-transition turns first order after the tricritical point. A pictorial depiction of the change in the potential landscape for this model can be found in Fig. 1 of Ref.~\cite{Lassig:1990xy}. The actual tricritical point for the quantum Hamiltonian is located numerically using DMRG~(see below).

In the Ginzburg-Landau formulation, the six primary fields of the tricritical Ising model can be identified with various~(normal ordered) powers of a field $\Phi$~[with Kac label (2,2)]~\cite{diFrancesco1997}. Two of the six fields are odd under the $\mathbb{Z}_2$ symmetry in the tricritical Ising model associated with the transformation $\Phi\rightarrow -\Phi$, while the others are even~(see Sec. 6.1 of  Ref.~\cite{Lencses:2023hlq} for a recent summary). For the quantum circuit model, this translates to the symmetry of the lattice operator~$\phi_j$ under the transformation $\phi_j\rightarrow -\phi_j$. Note that this is different from the $\mathbb{Z}_2$ symmetry in the Ising case~[see around Eq.~\eqref{eq:O_def}]. In the current model, the $\mathbb{Z}_2$-symmetry operator is simply the charge conjugation operator~[Eq.~\eqref{eq:C_act}]. 

The lattice operators corresponding to the two $\mathbb{Z}_2$-odd fields of the tricritical Ising model are given by
\begin{align}
\label{eq:s_tci}
\sigma \sim \sum_{k = 1, 2, \ldots} c_k\sin(k\phi_j) + \ldots,\nonumber \\\sigma' \sim \sum_{k = 1,2,\ldots} c'_k\sin(k\phi_j) + \ldots,
\end{align}
while the same for the~$\mathbb{Z}_2$-even fields are 
\begin{align}
\label{eq:e_tci}
\varepsilon \sim \sum_{k = 1, 2, \ldots} d_k\cos(k\phi_j) + \ldots,\nonumber \\\varepsilon' \sim \sum_{k = 1,2,\ldots} d'_k\cos(k\phi_j) + \ldots,\nonumber\\\varepsilon'' \sim \sum_{k = 1,2,\ldots} d''_k\cos(k\phi_j) + \ldots.
\end{align}
Here $c_k, c'_k,$ $d_k, d'_k, d''_k$ are non-universal lattice-dependent coefficients and the dots indicate subleading contributions.
\begin{figure}
    \centering
    \includegraphics[width=0.5\textwidth]{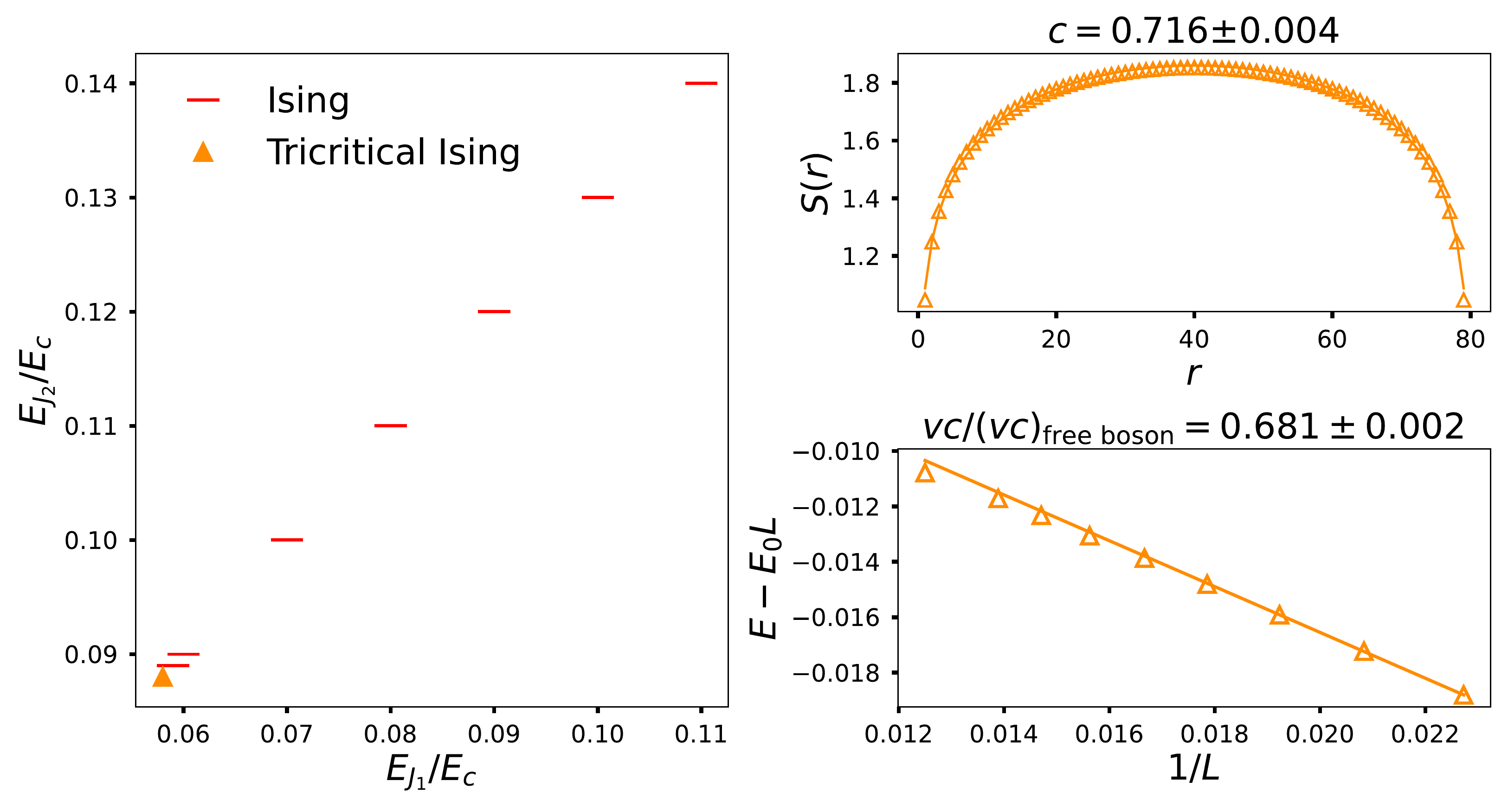}
    \caption{ \label{fig:fig_tci} DMRG results for the tricritical Ising point. The sine-Gordon coupling was chosen to be~$\beta^2\approx0.376$. (Left) The tricritical Ising point~(filled triangle) located at the end of an Ising critical line~(horizontal line markers). The critical points are located by computing the scaling of the entanglement entropy~$S$ as a function of the subsystem size~$r$~[Eq.~\eqref{eq:S_r}]. (Top right) Scaling of the entanglement entropy at the tricritical point for a system size $L=80$. The obtained value of the central charge is close to the expected value of 0.7 for the tricritical Ising field theory. Similar to the Ising case~(see Fig.~\ref{fig:Ising_1}), the discrepancy is due to the finite size effect. This was checked by computing the central charge for $L = 44$ to $80$ in steps of 2. (Bottom right) The scaling of the Casimir energy as a function of $1/L$~[see Eq.~\eqref{eq:cas}]. The obtained central charge is close to the expected value. The discrepancy occurs due to the slow convergence of energy with increasing bond-dimension. Note that the precision of the simulation is lower than the Ising case due to the larger local Hilbert space dimension~(27 instead of 17). The critical points were located with an accuracy of 0.002~(the Ising markers should be not confused with error bars).} 
\end{figure}

The DMRG analysis for this model is computationally more challenging due to the larger local Hilbert space dimension that needs to be manipulated to avoid truncation errors. In contrast to the Ising case, the local Hilbert space dimension was truncated to 27 and system sizes between 44 and 80 were simulated. Fig.~\ref{fig:fig_tci} presents the results obtained using DMRG for $\beta^2\approx0.376$~(similar results were obtained for $\beta^2\approx0.481$ which are not shown for brevity). The left panel shows the location of the Ising phase-transition line~(red dashes). The orange triangle indicates the location of the tricritical point. The central charge at the critical point was computed in two ways. First, the scaling of the entanglement entropy~[Eq.~\eqref{eq:S_r}] is shown on the top right panel for $L = 80$. The obtained result is close to the expected value of 7/10. The discrepancy is due to finite size effect. This was verified by simulating system sizes between~$L = 44$ to $80$ in steps of 2. Second, the scaling of the Casimir energy with system size~[Eq.~\eqref{eq:cas}] is computed and compared with that for the free-boson. As in the Ising case, the ratio of the product $vc$ for the tricritical Ising to the free boson case should yield the central charge of the tricritical Ising model. This is computed to be $\approx 0.681$, which is close to the expected value. The discrepancy is due to the slow convergence of the energy with increasing bond-dimension. With larger scale computations, the precision of these computations could be improved. The Ising transition line continues further as $E_{J_2}/E_c$ is increased, but the numerical simulations were restricted to the region shown in Fig.~\ref{fig:fig_tci}. 

The correlation functions of the primary fields can be verified by choosing the coefficients in Eqs.~(\ref{eq:s_tci}, \ref{eq:e_tci}) such that the resultant correlation function yields a scaling exponent that is close to the predictions for the model. Notice that the renormalization group flow from the tricritical Ising to the Ising critical point~(left panel of Fig.~\ref{fig:fig_tci}) is induced by changing the couplings~$E_{J_1}$ and $E_{J_2}$.  The corresponds to perturbing the lattice Hamiltonian at the tricritical point by a superposition of operators~$\cos\phi_j$ and~$\cos2\phi_j$. This is compatible with the identification of the field~$\varepsilon'$~[with Kac label~(1,3)] in Eq.~\eqref{eq:e_tci}, which is known to induce this flow. Finally, we note that in addition to the listed primary fields, the tricritical Ising model contains supersymmetric fields. The latter could be constructed for the quantum circuit by considering the fermionic operators built out of $\re^{\ri\alpha_1\pi n_j}, \re^{\ri\alpha_2\phi_k}$ by appropriately choosing $\alpha_1, \alpha_2$. We leave a more detailed analysis of the supersymmetric fields for a later work.

\section{Summary and Outlook}
\label{sec:concl}
In summary, a set of QEC lattices are described to realize multicritical Ising models in two space-time dimensions. The QEC lattices are based on circuit elements which are generalizations of ordinary Josephson junctions and give rise to potentials of the form $\cos(p\phi)$,~$p\in\mathbb{N}$. The elements for~$p = 1,2$ are well-known with the elements for~$p>2$ being straightforwardly realizable using recursive application of the scheme of Ref.~\cite{Brooks2013} with the elements for~$p = 1,2$. Starting with the QEC realization of the quantum sine-Gordon model with~$p$-fold degenerate minima, systematic perturbations are constructed using QEC elements to give rise to the multicritical Ising models. The cases of the Ising and the tricritical Ising models were analyzed. The next model in the series is the tetracritical Ising model corresponding to $p = 4$. The corresponding phases should be chosen as:
\begin{align}
\delta_1 = -\frac{\pi}{4}, \delta_2 = \frac{\pi}{2}, \delta_3 = -\frac{3\pi}{4}.
\end{align}
The location of the tetracritical point can, in principle, be obtained by tuning the three couplings:~$E_{J_n}$,~$n = 1,2,3$. 

In this way, quantum circuits can be used to systematically probe bulk and boundary perturbations of the multicritical Ising models venturing beyond the usually analyzed case of perturbed free boson theory~\cite{Caldeira1983, Schmid1983}~(see also Ref.~\cite{Houzet2020} for a recent work). Further generalizations can give rise to topological/perfectly-transmissive defects of conformal field theories~\cite{Petkova:2000ip, Bachas2001, Frohlich2004, Frohlich2006}. These defects commute with the generators of the conformal transformations and are deeply intertwined with the symmetries of the theory; for the Ising case, see Ref.~\cite{Grimm2001} for the spectrum of the lattice Hamiltonian and Refs.~\cite{Roy2021a, Rogerson:2022yim} for the entanglement properties of the ground state. Despite their importance in conformal field theories and 2+1D topological quantum field theories~\cite{Buican2017}, no systematic scheme is currently available for realizing these topological defects in a physical system. Their realization with QECs together with computation of transport  signatures of the topological defects would serve as a crucial step towards solving this problem. Finally, stacks of multicritical Ising chains with appropriate couplings can give rise to topologically ordered phases~\cite{Teo2014, Mong2014} which are a precious resource for the realization of topological quantum computation~\cite{Kitaev2003}. This amounts to stacking the one-dimensional chains of Fig.~\ref{fig:schematic} with suitable interactions and can lead to a systematic scheme for realization of topological matter with quantum circuits. We hope to return to some of these questions in the future.

Before concluding, we note that the current experimental works~\cite{Kuzmin2019, Leger2019}~(see Refs.~\cite{Houzet2019, Houzet2021} for related theoretical works) have been performed on systems where disorder plays a dominant role. It is conceivable that engineering a clean enough system with a suitable number of Josephson junctions will permit investigation of the multicritical phenomena described in this work~(see Ref.~\cite{Bell2018} for a related experimental work).
\section*{Acknowledgements}
Discussions with Michael Levin, Sergei Lukyanov and Hubert Saleur are gratefully acknowledged. AR was supported by a grant from the Simons Foundation (825876, TDN).

\bibliography{/Users/ananda/Dropbox/Bibliography/library_1}

\begin{thebibliography}{83}%
\makeatletter
\providecommand \@ifxundefined [1]{%
 \@ifx{#1\undefined}
}%
\providecommand \@ifnum [1]{%
 \ifnum #1\expandafter \@firstoftwo
 \else \expandafter \@secondoftwo
 \fi
}%
\providecommand \@ifx [1]{%
 \ifx #1\expandafter \@firstoftwo
 \else \expandafter \@secondoftwo
 \fi
}%
\providecommand \natexlab [1]{#1}%
\providecommand \enquote  [1]{``#1''}%
\providecommand \bibnamefont  [1]{#1}%
\providecommand \bibfnamefont [1]{#1}%
\providecommand \citenamefont [1]{#1}%
\providecommand \href@noop [0]{\@secondoftwo}%
\providecommand \href [0]{\begingroup \@sanitize@url \@href}%
\providecommand \@href[1]{\@@startlink{#1}\@@href}%
\providecommand \@@href[1]{\endgroup#1\@@endlink}%
\providecommand \@sanitize@url [0]{\catcode `\\12\catcode `\$12\catcode
  `\&12\catcode `\#12\catcode `\^12\catcode `\_12\catcode `\%12\relax}%
\providecommand \@@startlink[1]{}%
\providecommand \@@endlink[0]{}%
\providecommand \url  [0]{\begingroup\@sanitize@url \@url }%
\providecommand \@url [1]{\endgroup\@href {#1}{\urlprefix }}%
\providecommand \urlprefix  [0]{URL }%
\providecommand \Eprint [0]{\href }%
\providecommand \doibase [0]{https://doi.org/}%
\providecommand \selectlanguage [0]{\@gobble}%
\providecommand \bibinfo  [0]{\@secondoftwo}%
\providecommand \bibfield  [0]{\@secondoftwo}%
\providecommand \translation [1]{[#1]}%
\providecommand \BibitemOpen [0]{}%
\providecommand \bibitemStop [0]{}%
\providecommand \bibitemNoStop [0]{.\EOS\space}%
\providecommand \EOS [0]{\spacefactor3000\relax}%
\providecommand \BibitemShut  [1]{\csname bibitem#1\endcsname}%
\let\auto@bib@innerbib\@empty
\bibitem [{\citenamefont {Feynman}(1982)}]{Feynman_1982}%
  \BibitemOpen
  \bibfield  {author} {\bibinfo {author} {\bibfnamefont {R.~P.}\ \bibnamefont
  {Feynman}},\ }\bibfield  {title} {\bibinfo {title} {{Simulating Physics with
  Quantum Computers}},\ }\href@noop {} {\bibfield  {journal} {\bibinfo
  {journal} {Inernational Journal of Theoretical Physics}\ }\textbf {\bibinfo
  {volume} {21}},\ \bibinfo {pages} {467} (\bibinfo {year} {1982})}\BibitemShut
  {NoStop}%
\bibitem [{\citenamefont {Abrams}\ and\ \citenamefont
  {Lloyd}(1997)}]{Lloyd1997}%
  \BibitemOpen
  \bibfield  {author} {\bibinfo {author} {\bibfnamefont {D.~S.}\ \bibnamefont
  {Abrams}}\ and\ \bibinfo {author} {\bibfnamefont {S.}~\bibnamefont {Lloyd}},\
  }\bibfield  {title} {\bibinfo {title} {Simulation of many-body fermi systems
  on a universal quantum computer},\ }\href
  {https://doi.org/10.1103/PhysRevLett.79.2586} {\bibfield  {journal} {\bibinfo
   {journal} {Phys. Rev. Lett.}\ }\textbf {\bibinfo {volume} {79}},\ \bibinfo
  {pages} {2586} (\bibinfo {year} {1997})}\BibitemShut {NoStop}%
\bibitem [{\citenamefont {Jordan}\ \emph {et~al.}(2014)\citenamefont {Jordan},
  \citenamefont {Lee},\ and\ \citenamefont {Preskill}}]{Jordan2011}%
  \BibitemOpen
  \bibfield  {author} {\bibinfo {author} {\bibfnamefont {S.~P.}\ \bibnamefont
  {Jordan}}, \bibinfo {author} {\bibfnamefont {K.~S.~M.}\ \bibnamefont {Lee}},\
  and\ \bibinfo {author} {\bibfnamefont {J.}~\bibnamefont {Preskill}},\
  }\bibfield  {title} {\bibinfo {title} {{Quantum Computation of Scattering in
  Scalar Quantum Field Theories}},\ }\href@noop {} {\bibfield  {journal}
  {\bibinfo  {journal} {Quant. Inf. Comput.}\ }\textbf {\bibinfo {volume} {14}}
  (\bibinfo {year} {2014})},\ \Eprint {https://arxiv.org/abs/1112.4833}
  {arXiv:1112.4833 [hep-th]} \BibitemShut {NoStop}%
\bibitem [{\citenamefont {Jordan}\ \emph {et~al.}(2012)\citenamefont {Jordan},
  \citenamefont {Lee},\ and\ \citenamefont {Preskill}}]{Jordan2012}%
  \BibitemOpen
  \bibfield  {author} {\bibinfo {author} {\bibfnamefont {S.~P.}\ \bibnamefont
  {Jordan}}, \bibinfo {author} {\bibfnamefont {K.~S.~M.}\ \bibnamefont {Lee}},\
  and\ \bibinfo {author} {\bibfnamefont {J.}~\bibnamefont {Preskill}},\
  }\bibfield  {title} {\bibinfo {title} {{Quantum Algorithms for Quantum Field
  Theories}},\ }\href {https://doi.org/10.1126/science.1217069} {\bibfield
  {journal} {\bibinfo  {journal} {Science}\ }\textbf {\bibinfo {volume}
  {336}},\ \bibinfo {pages} {1130} (\bibinfo {year} {2012})},\ \Eprint
  {https://arxiv.org/abs/1111.3633} {arXiv:1111.3633 [quant-ph]} \BibitemShut
  {NoStop}%
\bibitem [{\citenamefont {Smith}\ \emph {et~al.}(2019)\citenamefont {Smith},
  \citenamefont {Kim}, \citenamefont {Pollmann},\ and\ \citenamefont
  {Knolle}}]{Smith2019}%
  \BibitemOpen
  \bibfield  {author} {\bibinfo {author} {\bibfnamefont {A.}~\bibnamefont
  {Smith}}, \bibinfo {author} {\bibfnamefont {M.~S.}\ \bibnamefont {Kim}},
  \bibinfo {author} {\bibfnamefont {F.}~\bibnamefont {Pollmann}},\ and\
  \bibinfo {author} {\bibfnamefont {J.}~\bibnamefont {Knolle}},\ }\bibfield
  {title} {\bibinfo {title} {{Simulating quantum many-body dynamics on a
  current digital quantum computer}},\ }\href
  {https://doi.org/10.1038/s41534-019-0217-0} {\bibfield  {journal} {\bibinfo
  {journal} {npj Quantum Inf.}\ }\textbf {\bibinfo {volume} {5}},\ \bibinfo
  {pages} {106} (\bibinfo {year} {2019})},\ \Eprint
  {https://arxiv.org/abs/1906.06343} {arXiv:1906.06343 [quant-ph]} \BibitemShut
  {NoStop}%
\bibitem [{\citenamefont {Smith}\ \emph {et~al.}(2022)\citenamefont {Smith},
  \citenamefont {Jobst}, \citenamefont {Green},\ and\ \citenamefont
  {Pollmann}}]{Smith2022}%
  \BibitemOpen
  \bibfield  {author} {\bibinfo {author} {\bibfnamefont {A.}~\bibnamefont
  {Smith}}, \bibinfo {author} {\bibfnamefont {B.}~\bibnamefont {Jobst}},
  \bibinfo {author} {\bibfnamefont {A.~G.}\ \bibnamefont {Green}},\ and\
  \bibinfo {author} {\bibfnamefont {F.}~\bibnamefont {Pollmann}},\ }\bibfield
  {title} {\bibinfo {title} {Crossing a topological phase transition with a
  quantum computer},\ }\href
  {https://doi.org/10.1103/PhysRevResearch.4.L022020} {\bibfield  {journal}
  {\bibinfo  {journal} {Phys. Rev. Res.}\ }\textbf {\bibinfo {volume} {4}},\
  \bibinfo {pages} {L022020} (\bibinfo {year} {2022})}\BibitemShut {NoStop}%
\bibitem [{\citenamefont {Vovrosh}\ and\ \citenamefont
  {Knolle}(2021)}]{Vovrosh2021}%
  \BibitemOpen
  \bibfield  {author} {\bibinfo {author} {\bibfnamefont {J.}~\bibnamefont
  {Vovrosh}}\ and\ \bibinfo {author} {\bibfnamefont {J.}~\bibnamefont
  {Knolle}},\ }\bibfield  {title} {\bibinfo {title} {Confinement and
  entanglement dynamics on a digital quantum computer},\ }\href
  {https://doi.org/10.1038/s41598-021-90849-5} {\bibfield  {journal} {\bibinfo
  {journal} {Scientific Reports}\ }\textbf {\bibinfo {volume} {11}},\ \bibinfo
  {pages} {11577} (\bibinfo {year} {2021})}\BibitemShut {NoStop}%
\bibitem [{\citenamefont {Tan}\ \emph {et~al.}(2021)\citenamefont {Tan},
  \citenamefont {Becker}, \citenamefont {Liu}, \citenamefont {Pagano},
  \citenamefont {Collins}, \citenamefont {De}, \citenamefont {Feng},
  \citenamefont {Kaplan}, \citenamefont {Kyprianidis}, \citenamefont
  {Lundgren}, \citenamefont {Morong}, \citenamefont {Whitsitt}, \citenamefont
  {Gorshkov},\ and\ \citenamefont {Monroe}}]{Tan_2021}%
  \BibitemOpen
  \bibfield  {author} {\bibinfo {author} {\bibfnamefont {W.~L.}\ \bibnamefont
  {Tan}}, \bibinfo {author} {\bibfnamefont {P.}~\bibnamefont {Becker}},
  \bibinfo {author} {\bibfnamefont {F.}~\bibnamefont {Liu}}, \bibinfo {author}
  {\bibfnamefont {G.}~\bibnamefont {Pagano}}, \bibinfo {author} {\bibfnamefont
  {K.~S.}\ \bibnamefont {Collins}}, \bibinfo {author} {\bibfnamefont
  {A.}~\bibnamefont {De}}, \bibinfo {author} {\bibfnamefont {L.}~\bibnamefont
  {Feng}}, \bibinfo {author} {\bibfnamefont {H.~B.}\ \bibnamefont {Kaplan}},
  \bibinfo {author} {\bibfnamefont {A.}~\bibnamefont {Kyprianidis}}, \bibinfo
  {author} {\bibfnamefont {R.}~\bibnamefont {Lundgren}}, \bibinfo {author}
  {\bibfnamefont {W.}~\bibnamefont {Morong}}, \bibinfo {author} {\bibfnamefont
  {S.}~\bibnamefont {Whitsitt}}, \bibinfo {author} {\bibfnamefont {A.~V.}\
  \bibnamefont {Gorshkov}},\ and\ \bibinfo {author} {\bibfnamefont
  {C.}~\bibnamefont {Monroe}},\ }\bibfield  {title} {\bibinfo {title}
  {Domain-wall confinement and dynamics in a quantum simulator},\ }\href
  {https://doi.org/10.1038/s41567-021-01194-3} {\bibfield  {journal} {\bibinfo
  {journal} {Nature Physics}\ }\textbf {\bibinfo {volume} {17}},\ \bibinfo
  {pages} {742} (\bibinfo {year} {2021})}\BibitemShut {NoStop}%
\bibitem [{\citenamefont {{Lamb}}\ \emph {et~al.}(2023)\citenamefont {{Lamb}},
  \citenamefont {{Tang}}, \citenamefont {{Davis}},\ and\ \citenamefont
  {{Roy}}}]{Lamb2023}%
  \BibitemOpen
  \bibfield  {author} {\bibinfo {author} {\bibfnamefont {C.}~\bibnamefont
  {{Lamb}}}, \bibinfo {author} {\bibfnamefont {Y.}~\bibnamefont {{Tang}}},
  \bibinfo {author} {\bibfnamefont {R.}~\bibnamefont {{Davis}}},\ and\ \bibinfo
  {author} {\bibfnamefont {A.}~\bibnamefont {{Roy}}},\ }\bibfield  {title}
  {\bibinfo {title} {{Ising Meson Spectroscopy on a Noisy Digital Quantum
  Simulator}},\ }\href {https://doi.org/10.48550/arXiv.2303.03311} {\bibfield
  {journal} {\bibinfo  {journal} {arXiv e-prints}\ ,\ \bibinfo {eid}
  {arXiv:2303.03311}} (\bibinfo {year} {2023})},\ \Eprint
  {https://arxiv.org/abs/2303.03311} {arXiv:2303.03311 [quant-ph]} \BibitemShut
  {NoStop}%
\bibitem [{\citenamefont {Dou\ifmmode~\mbox{\c{c}}\else \c{c}\fi{}ot}\ \emph
  {et~al.}(2004)\citenamefont {Dou\ifmmode~\mbox{\c{c}}\else \c{c}\fi{}ot},
  \citenamefont {Ioffe},\ and\ \citenamefont {Vidal}}]{Doucot2004}%
  \BibitemOpen
  \bibfield  {author} {\bibinfo {author} {\bibfnamefont {B.}~\bibnamefont
  {Dou\ifmmode~\mbox{\c{c}}\else \c{c}\fi{}ot}}, \bibinfo {author}
  {\bibfnamefont {L.~B.}\ \bibnamefont {Ioffe}},\ and\ \bibinfo {author}
  {\bibfnamefont {J.}~\bibnamefont {Vidal}},\ }\bibfield  {title} {\bibinfo
  {title} {Discrete non-abelian gauge theories in josephson-junction arrays and
  quantum computation},\ }\href {https://doi.org/10.1103/PhysRevB.69.214501}
  {\bibfield  {journal} {\bibinfo  {journal} {Phys. Rev. B}\ }\textbf {\bibinfo
  {volume} {69}},\ \bibinfo {pages} {214501} (\bibinfo {year}
  {2004})}\BibitemShut {NoStop}%
\bibitem [{\citenamefont {Cirac}\ \emph {et~al.}(2010)\citenamefont {Cirac},
  \citenamefont {Maraner},\ and\ \citenamefont {Pachos}}]{Cirac2010}%
  \BibitemOpen
  \bibfield  {author} {\bibinfo {author} {\bibfnamefont {J.~I.}\ \bibnamefont
  {Cirac}}, \bibinfo {author} {\bibfnamefont {P.}~\bibnamefont {Maraner}},\
  and\ \bibinfo {author} {\bibfnamefont {J.~K.}\ \bibnamefont {Pachos}},\
  }\bibfield  {title} {\bibinfo {title} {Cold atom simulation of interacting
  relativistic quantum field theories},\ }\href
  {https://doi.org/10.1103/PhysRevLett.105.190403} {\bibfield  {journal}
  {\bibinfo  {journal} {Phys. Rev. Lett.}\ }\textbf {\bibinfo {volume} {105}},\
  \bibinfo {pages} {190403} (\bibinfo {year} {2010})}\BibitemShut {NoStop}%
\bibitem [{\citenamefont {B\"uchler}\ \emph {et~al.}(2005)\citenamefont
  {B\"uchler}, \citenamefont {Hermele}, \citenamefont {Huber}, \citenamefont
  {Fisher},\ and\ \citenamefont {Zoller}}]{Buchler2005}%
  \BibitemOpen
  \bibfield  {author} {\bibinfo {author} {\bibfnamefont {H.~P.}\ \bibnamefont
  {B\"uchler}}, \bibinfo {author} {\bibfnamefont {M.}~\bibnamefont {Hermele}},
  \bibinfo {author} {\bibfnamefont {S.~D.}\ \bibnamefont {Huber}}, \bibinfo
  {author} {\bibfnamefont {M.~P.~A.}\ \bibnamefont {Fisher}},\ and\ \bibinfo
  {author} {\bibfnamefont {P.}~\bibnamefont {Zoller}},\ }\bibfield  {title}
  {\bibinfo {title} {Atomic quantum simulator for lattice gauge theories and
  ring exchange models},\ }\href
  {https://doi.org/10.1103/PhysRevLett.95.040402} {\bibfield  {journal}
  {\bibinfo  {journal} {Phys. Rev. Lett.}\ }\textbf {\bibinfo {volume} {95}},\
  \bibinfo {pages} {040402} (\bibinfo {year} {2005})}\BibitemShut {NoStop}%
\bibitem [{\citenamefont {Casanova}\ \emph {et~al.}(2011)\citenamefont
  {Casanova}, \citenamefont {Lamata}, \citenamefont {Egusquiza}, \citenamefont
  {Gerritsma}, \citenamefont {Roos}, \citenamefont {Garc\'{\i}a-Ripoll},\ and\
  \citenamefont {Solano}}]{Casanova2011}%
  \BibitemOpen
  \bibfield  {author} {\bibinfo {author} {\bibfnamefont {J.}~\bibnamefont
  {Casanova}}, \bibinfo {author} {\bibfnamefont {L.}~\bibnamefont {Lamata}},
  \bibinfo {author} {\bibfnamefont {I.~L.}\ \bibnamefont {Egusquiza}}, \bibinfo
  {author} {\bibfnamefont {R.}~\bibnamefont {Gerritsma}}, \bibinfo {author}
  {\bibfnamefont {C.~F.}\ \bibnamefont {Roos}}, \bibinfo {author}
  {\bibfnamefont {J.~J.}\ \bibnamefont {Garc\'{\i}a-Ripoll}},\ and\ \bibinfo
  {author} {\bibfnamefont {E.}~\bibnamefont {Solano}},\ }\bibfield  {title}
  {\bibinfo {title} {Quantum simulation of quantum field theories in trapped
  ions},\ }\href {https://doi.org/10.1103/PhysRevLett.107.260501} {\bibfield
  {journal} {\bibinfo  {journal} {Phys. Rev. Lett.}\ }\textbf {\bibinfo
  {volume} {107}},\ \bibinfo {pages} {260501} (\bibinfo {year}
  {2011})}\BibitemShut {NoStop}%
\bibitem [{\citenamefont {Roy}\ and\ \citenamefont {Saleur}(2019)}]{Roy2019}%
  \BibitemOpen
  \bibfield  {author} {\bibinfo {author} {\bibfnamefont {A.}~\bibnamefont
  {Roy}}\ and\ \bibinfo {author} {\bibfnamefont {H.}~\bibnamefont {Saleur}},\
  }\bibfield  {title} {\bibinfo {title} {Quantum electronic circuit simulation
  of generalized sine-gordon models},\ }\href
  {https://doi.org/10.1103/PhysRevB.100.155425} {\bibfield  {journal} {\bibinfo
   {journal} {Phys. Rev. B}\ }\textbf {\bibinfo {volume} {100}},\ \bibinfo
  {pages} {155425} (\bibinfo {year} {2019})}\BibitemShut {NoStop}%
\bibitem [{\citenamefont {Duca}\ \emph {et~al.}(2015)\citenamefont {Duca},
  \citenamefont {Li}, \citenamefont {Reitter}, \citenamefont {Bloch},
  \citenamefont {Schleier-Smith},\ and\ \citenamefont {Schneider}}]{Duca2015}%
  \BibitemOpen
  \bibfield  {author} {\bibinfo {author} {\bibfnamefont {L.}~\bibnamefont
  {Duca}}, \bibinfo {author} {\bibfnamefont {T.}~\bibnamefont {Li}}, \bibinfo
  {author} {\bibfnamefont {M.}~\bibnamefont {Reitter}}, \bibinfo {author}
  {\bibfnamefont {I.}~\bibnamefont {Bloch}}, \bibinfo {author} {\bibfnamefont
  {M.}~\bibnamefont {Schleier-Smith}},\ and\ \bibinfo {author} {\bibfnamefont
  {U.}~\bibnamefont {Schneider}},\ }\bibfield  {title} {\bibinfo {title} {An
  aharonov-bohm interferometer for determining bloch band topology},\ }\href
  {https://doi.org/10.1126/science.1259052} {\bibfield  {journal} {\bibinfo
  {journal} {Science}\ }\textbf {\bibinfo {volume} {347}},\ \bibinfo {pages}
  {288} (\bibinfo {year} {2015})},\ \Eprint
  {https://arxiv.org/abs/https://science.sciencemag.org/content/347/6219/288.full.pdf}
  {https://science.sciencemag.org/content/347/6219/288.full.pdf} \BibitemShut
  {NoStop}%
\bibitem [{\citenamefont {Parsons}\ \emph {et~al.}(2016)\citenamefont
  {Parsons}, \citenamefont {Mazurenko}, \citenamefont {Chiu}, \citenamefont
  {Ji}, \citenamefont {Greif},\ and\ \citenamefont {Greiner}}]{Parsons2016}%
  \BibitemOpen
  \bibfield  {author} {\bibinfo {author} {\bibfnamefont {M.~F.}\ \bibnamefont
  {Parsons}}, \bibinfo {author} {\bibfnamefont {A.}~\bibnamefont {Mazurenko}},
  \bibinfo {author} {\bibfnamefont {C.~S.}\ \bibnamefont {Chiu}}, \bibinfo
  {author} {\bibfnamefont {G.}~\bibnamefont {Ji}}, \bibinfo {author}
  {\bibfnamefont {D.}~\bibnamefont {Greif}},\ and\ \bibinfo {author}
  {\bibfnamefont {M.}~\bibnamefont {Greiner}},\ }\bibfield  {title} {\bibinfo
  {title} {Site-resolved measurement of the spin-correlation function in the
  fermi-hubbard model},\ }\href {https://doi.org/10.1126/science.aag1430}
  {\bibfield  {journal} {\bibinfo  {journal} {Science}\ }\textbf {\bibinfo
  {volume} {353}},\ \bibinfo {pages} {1253} (\bibinfo {year} {2016})},\ \Eprint
  {https://arxiv.org/abs/https://science.sciencemag.org/content/353/6305/1253.full.pdf}
  {https://science.sciencemag.org/content/353/6305/1253.full.pdf} \BibitemShut
  {NoStop}%
\bibitem [{\citenamefont {Gross}\ and\ \citenamefont
  {Bloch}(2017)}]{Gross2017}%
  \BibitemOpen
  \bibfield  {author} {\bibinfo {author} {\bibfnamefont {C.}~\bibnamefont
  {Gross}}\ and\ \bibinfo {author} {\bibfnamefont {I.}~\bibnamefont {Bloch}},\
  }\bibfield  {title} {\bibinfo {title} {Quantum simulations with ultracold
  atoms in optical lattices},\ }\href {https://doi.org/10.1126/science.aal3837}
  {\bibfield  {journal} {\bibinfo  {journal} {Science}\ }\textbf {\bibinfo
  {volume} {357}},\ \bibinfo {pages} {995} (\bibinfo {year} {2017})},\ \Eprint
  {https://arxiv.org/abs/https://science.sciencemag.org/content/357/6355/995.full.pdf}
  {https://science.sciencemag.org/content/357/6355/995.full.pdf} \BibitemShut
  {NoStop}%
\bibitem [{\citenamefont {{Bernien}}\ \emph {et~al.}(2017)\citenamefont
  {{Bernien}}, \citenamefont {{Schwartz}}, \citenamefont {{Keesling}},
  \citenamefont {{Levine}}, \citenamefont {{Omran}}, \citenamefont {{Pichler}},
  \citenamefont {{Choi}}, \citenamefont {{Zibrov}}, \citenamefont {{Endres}},
  \citenamefont {{Greiner}}, \citenamefont {{Vuleti{\'c}}},\ and\ \citenamefont
  {{Lukin}}}]{Bernien2017}%
  \BibitemOpen
  \bibfield  {author} {\bibinfo {author} {\bibfnamefont {H.}~\bibnamefont
  {{Bernien}}}, \bibinfo {author} {\bibfnamefont {S.}~\bibnamefont
  {{Schwartz}}}, \bibinfo {author} {\bibfnamefont {A.}~\bibnamefont
  {{Keesling}}}, \bibinfo {author} {\bibfnamefont {H.}~\bibnamefont
  {{Levine}}}, \bibinfo {author} {\bibfnamefont {A.}~\bibnamefont {{Omran}}},
  \bibinfo {author} {\bibfnamefont {H.}~\bibnamefont {{Pichler}}}, \bibinfo
  {author} {\bibfnamefont {S.}~\bibnamefont {{Choi}}}, \bibinfo {author}
  {\bibfnamefont {A.~S.}\ \bibnamefont {{Zibrov}}}, \bibinfo {author}
  {\bibfnamefont {M.}~\bibnamefont {{Endres}}}, \bibinfo {author}
  {\bibfnamefont {M.}~\bibnamefont {{Greiner}}}, \bibinfo {author}
  {\bibfnamefont {V.}~\bibnamefont {{Vuleti{\'c}}}},\ and\ \bibinfo {author}
  {\bibfnamefont {M.~D.}\ \bibnamefont {{Lukin}}},\ }\bibfield  {title}
  {\bibinfo {title} {{Probing many-body dynamics on a 51-atom quantum
  simulator}},\ }\href {https://doi.org/10.1038/nature24622} {\bibfield
  {journal} {\bibinfo  {journal} {Nature}\ }\textbf {\bibinfo {volume} {551}},\
  \bibinfo {pages} {579} (\bibinfo {year} {2017})},\ \Eprint
  {https://arxiv.org/abs/1707.04344} {arXiv:1707.04344 [quant-ph]} \BibitemShut
  {NoStop}%
\bibitem [{\citenamefont {Samajdar}\ \emph {et~al.}(2020)\citenamefont
  {Samajdar}, \citenamefont {Ho}, \citenamefont {Pichler}, \citenamefont
  {Lukin},\ and\ \citenamefont {Sachdev}}]{Samajdar2020}%
  \BibitemOpen
  \bibfield  {author} {\bibinfo {author} {\bibfnamefont {R.}~\bibnamefont
  {Samajdar}}, \bibinfo {author} {\bibfnamefont {W.~W.}\ \bibnamefont {Ho}},
  \bibinfo {author} {\bibfnamefont {H.}~\bibnamefont {Pichler}}, \bibinfo
  {author} {\bibfnamefont {M.~D.}\ \bibnamefont {Lukin}},\ and\ \bibinfo
  {author} {\bibfnamefont {S.}~\bibnamefont {Sachdev}},\ }\bibfield  {title}
  {\bibinfo {title} {Quantum phases of rydberg atoms on a kagome lattice},\
  }\href@noop {} {\bibfield  {journal} {\bibinfo  {journal} {Proceedings of the
  National Academy of Sciences}\ }\textbf {\bibinfo {volume} {118}} (\bibinfo
  {year} {2020})}\BibitemShut {NoStop}%
\bibitem [{\citenamefont {Ebadi}\ \emph {et~al.}(2021)\citenamefont {Ebadi},
  \citenamefont {Wang}, \citenamefont {Levine}, \citenamefont {Keesling},
  \citenamefont {Semeghini}, \citenamefont {Omran}, \citenamefont {Bluvstein},
  \citenamefont {Samajdar}, \citenamefont {Pichler}, \citenamefont {Ho},
  \citenamefont {Choi}, \citenamefont {Sachdev}, \citenamefont {Greiner},
  \citenamefont {Vuleti{\'{c}}},\ and\ \citenamefont {Lukin}}]{Ebadi_2021}%
  \BibitemOpen
  \bibfield  {author} {\bibinfo {author} {\bibfnamefont {S.}~\bibnamefont
  {Ebadi}}, \bibinfo {author} {\bibfnamefont {T.~T.}\ \bibnamefont {Wang}},
  \bibinfo {author} {\bibfnamefont {H.}~\bibnamefont {Levine}}, \bibinfo
  {author} {\bibfnamefont {A.}~\bibnamefont {Keesling}}, \bibinfo {author}
  {\bibfnamefont {G.}~\bibnamefont {Semeghini}}, \bibinfo {author}
  {\bibfnamefont {A.}~\bibnamefont {Omran}}, \bibinfo {author} {\bibfnamefont
  {D.}~\bibnamefont {Bluvstein}}, \bibinfo {author} {\bibfnamefont
  {R.}~\bibnamefont {Samajdar}}, \bibinfo {author} {\bibfnamefont
  {H.}~\bibnamefont {Pichler}}, \bibinfo {author} {\bibfnamefont {W.~W.}\
  \bibnamefont {Ho}}, \bibinfo {author} {\bibfnamefont {S.}~\bibnamefont
  {Choi}}, \bibinfo {author} {\bibfnamefont {S.}~\bibnamefont {Sachdev}},
  \bibinfo {author} {\bibfnamefont {M.}~\bibnamefont {Greiner}}, \bibinfo
  {author} {\bibfnamefont {V.}~\bibnamefont {Vuleti{\'{c}}}},\ and\ \bibinfo
  {author} {\bibfnamefont {M.~D.}\ \bibnamefont {Lukin}},\ }\bibfield  {title}
  {\bibinfo {title} {Quantum phases of matter on a 256-atom programmable
  quantum simulator},\ }\href {https://doi.org/10.1038/s41586-021-03582-4}
  {\bibfield  {journal} {\bibinfo  {journal} {Nature}\ }\textbf {\bibinfo
  {volume} {595}},\ \bibinfo {pages} {227} (\bibinfo {year}
  {2021})}\BibitemShut {NoStop}%
\bibitem [{\citenamefont {Friedenauer}\ \emph {et~al.}(2008)\citenamefont
  {Friedenauer}, \citenamefont {Schmitz}, \citenamefont {Glueckert},
  \citenamefont {Porras},\ and\ \citenamefont {Schaetz}}]{Friedenauer2008}%
  \BibitemOpen
  \bibfield  {author} {\bibinfo {author} {\bibfnamefont {A.}~\bibnamefont
  {Friedenauer}}, \bibinfo {author} {\bibfnamefont {H.}~\bibnamefont
  {Schmitz}}, \bibinfo {author} {\bibfnamefont {J.~T.}\ \bibnamefont
  {Glueckert}}, \bibinfo {author} {\bibfnamefont {D.}~\bibnamefont {Porras}},\
  and\ \bibinfo {author} {\bibfnamefont {T.}~\bibnamefont {Schaetz}},\
  }\bibfield  {title} {\bibinfo {title} {Simulating a quantum magnet with
  trapped ions},\ }\href {https://doi.org/10.1038/nphys1032} {\bibfield
  {journal} {\bibinfo  {journal} {Nature Physics}\ }\textbf {\bibinfo {volume}
  {4}},\ \bibinfo {pages} {757} (\bibinfo {year} {2008})}\BibitemShut {NoStop}%
\bibitem [{\citenamefont {Kim}\ \emph {et~al.}(2010)\citenamefont {Kim},
  \citenamefont {Chang}, \citenamefont {Korenblit}, \citenamefont {Islam},
  \citenamefont {Edwards}, \citenamefont {Freericks}, \citenamefont {Lin},
  \citenamefont {Duan},\ and\ \citenamefont {Monroe}}]{Kim2010}%
  \BibitemOpen
  \bibfield  {author} {\bibinfo {author} {\bibfnamefont {K.}~\bibnamefont
  {Kim}}, \bibinfo {author} {\bibfnamefont {M.-S.}\ \bibnamefont {Chang}},
  \bibinfo {author} {\bibfnamefont {S.}~\bibnamefont {Korenblit}}, \bibinfo
  {author} {\bibfnamefont {R.}~\bibnamefont {Islam}}, \bibinfo {author}
  {\bibfnamefont {E.~E.}\ \bibnamefont {Edwards}}, \bibinfo {author}
  {\bibfnamefont {J.~K.}\ \bibnamefont {Freericks}}, \bibinfo {author}
  {\bibfnamefont {G.-D.}\ \bibnamefont {Lin}}, \bibinfo {author} {\bibfnamefont
  {L.-M.}\ \bibnamefont {Duan}},\ and\ \bibinfo {author} {\bibfnamefont
  {C.}~\bibnamefont {Monroe}},\ }\bibfield  {title} {\bibinfo {title} {Quantum
  simulation of frustrated ising spins with trapped ions},\ }\href
  {https://doi.org/10.1038/nature09071} {\bibfield  {journal} {\bibinfo
  {journal} {Nature}\ }\textbf {\bibinfo {volume} {465}},\ \bibinfo {pages}
  {590} (\bibinfo {year} {2010})}\BibitemShut {NoStop}%
\bibitem [{\citenamefont {Kuzmin}\ \emph
  {et~al.}(2019{\natexlab{a}})\citenamefont {Kuzmin}, \citenamefont {Mehta},
  \citenamefont {Grabon}, \citenamefont {Mencia},\ and\ \citenamefont
  {Manucharyan}}]{Kuzmin2018}%
  \BibitemOpen
  \bibfield  {author} {\bibinfo {author} {\bibfnamefont {R.}~\bibnamefont
  {Kuzmin}}, \bibinfo {author} {\bibfnamefont {N.}~\bibnamefont {Mehta}},
  \bibinfo {author} {\bibfnamefont {N.}~\bibnamefont {Grabon}}, \bibinfo
  {author} {\bibfnamefont {R.}~\bibnamefont {Mencia}},\ and\ \bibinfo {author}
  {\bibfnamefont {V.~E.}\ \bibnamefont {Manucharyan}},\ }\bibfield  {title}
  {\bibinfo {title} {Superstrong coupling in circuit quantum electrodynamics},\
  }\href {https://doi.org/10.1038/s41534-019-0134-2} {\bibfield  {journal}
  {\bibinfo  {journal} {NPJ Quantum Information}\ }\textbf {\bibinfo {volume}
  {5}},\ \bibinfo {pages} {20} (\bibinfo {year}
  {2019}{\natexlab{a}})}\BibitemShut {NoStop}%
\bibitem [{\citenamefont {Kuzmin}\ \emph
  {et~al.}(2019{\natexlab{b}})\citenamefont {Kuzmin}, \citenamefont {Mencia},
  \citenamefont {Grabon}, \citenamefont {Mehta}, \citenamefont {Lin},\ and\
  \citenamefont {Manucharyan}}]{Kuzmin2019}%
  \BibitemOpen
  \bibfield  {author} {\bibinfo {author} {\bibfnamefont {R.}~\bibnamefont
  {Kuzmin}}, \bibinfo {author} {\bibfnamefont {R.}~\bibnamefont {Mencia}},
  \bibinfo {author} {\bibfnamefont {N.}~\bibnamefont {Grabon}}, \bibinfo
  {author} {\bibfnamefont {N.}~\bibnamefont {Mehta}}, \bibinfo {author}
  {\bibfnamefont {Y.-H.}\ \bibnamefont {Lin}},\ and\ \bibinfo {author}
  {\bibfnamefont {V.~E.}\ \bibnamefont {Manucharyan}},\ }\bibfield  {title}
  {\bibinfo {title} {Quantum electrodynamics of a superconductor--insulator
  phase transition},\ }\href {https://doi.org/10.1038/s41567-019-0553-1}
  {\bibfield  {journal} {\bibinfo  {journal} {Nature Physics}\ }\textbf
  {\bibinfo {volume} {15}},\ \bibinfo {pages} {930} (\bibinfo {year}
  {2019}{\natexlab{b}})}\BibitemShut {NoStop}%
\bibitem [{\citenamefont {L{\'e}ger}\ \emph {et~al.}(2019)\citenamefont
  {L{\'e}ger}, \citenamefont {Puertas-Mart{\'i}nez}, \citenamefont {Bharadwaj},
  \citenamefont {Dassonneville}, \citenamefont {Delaforce}, \citenamefont
  {Foroughi}, \citenamefont {Milchakov}, \citenamefont {Planat}, \citenamefont
  {Buisson}, \citenamefont {Naud}, \citenamefont {Hasch-Guichard},
  \citenamefont {Florens}, \citenamefont {Snyman},\ and\ \citenamefont
  {Roch}}]{Leger2019}%
  \BibitemOpen
  \bibfield  {author} {\bibinfo {author} {\bibfnamefont {S.}~\bibnamefont
  {L{\'e}ger}}, \bibinfo {author} {\bibfnamefont {J.}~\bibnamefont
  {Puertas-Mart{\'i}nez}}, \bibinfo {author} {\bibfnamefont {K.}~\bibnamefont
  {Bharadwaj}}, \bibinfo {author} {\bibfnamefont {R.}~\bibnamefont
  {Dassonneville}}, \bibinfo {author} {\bibfnamefont {J.}~\bibnamefont
  {Delaforce}}, \bibinfo {author} {\bibfnamefont {F.}~\bibnamefont {Foroughi}},
  \bibinfo {author} {\bibfnamefont {V.}~\bibnamefont {Milchakov}}, \bibinfo
  {author} {\bibfnamefont {L.}~\bibnamefont {Planat}}, \bibinfo {author}
  {\bibfnamefont {O.}~\bibnamefont {Buisson}}, \bibinfo {author} {\bibfnamefont
  {C.}~\bibnamefont {Naud}}, \bibinfo {author} {\bibfnamefont {W.}~\bibnamefont
  {Hasch-Guichard}}, \bibinfo {author} {\bibfnamefont {S.}~\bibnamefont
  {Florens}}, \bibinfo {author} {\bibfnamefont {I.}~\bibnamefont {Snyman}},\
  and\ \bibinfo {author} {\bibfnamefont {N.}~\bibnamefont {Roch}},\ }\bibfield
  {title} {\bibinfo {title} {Observation of quantum many-body effects due to
  zero point fluctuations in superconducting circuits},\ }\href
  {https://doi.org/10.1038/s41467-019-13199-x} {\bibfield  {journal} {\bibinfo
  {journal} {Nature Communications}\ }\textbf {\bibinfo {volume} {10}},\
  \bibinfo {pages} {5259} (\bibinfo {year} {2019})}\BibitemShut {NoStop}%
\bibitem [{\citenamefont {Puertas~Mart{\'i}nez}\ \emph
  {et~al.}(2019)\citenamefont {Puertas~Mart{\'i}nez}, \citenamefont
  {L{\'e}ger}, \citenamefont {Gheeraert}, \citenamefont {Dassonneville},
  \citenamefont {Planat}, \citenamefont {Foroughi}, \citenamefont {Krupko},
  \citenamefont {Buisson}, \citenamefont {Naud}, \citenamefont
  {Hasch-Guichard}, \citenamefont {Florens}, \citenamefont {Snyman},\ and\
  \citenamefont {Roch}}]{PuertasMartinez2019}%
  \BibitemOpen
  \bibfield  {author} {\bibinfo {author} {\bibfnamefont {J.}~\bibnamefont
  {Puertas~Mart{\'i}nez}}, \bibinfo {author} {\bibfnamefont {S.}~\bibnamefont
  {L{\'e}ger}}, \bibinfo {author} {\bibfnamefont {N.}~\bibnamefont
  {Gheeraert}}, \bibinfo {author} {\bibfnamefont {R.}~\bibnamefont
  {Dassonneville}}, \bibinfo {author} {\bibfnamefont {L.}~\bibnamefont
  {Planat}}, \bibinfo {author} {\bibfnamefont {F.}~\bibnamefont {Foroughi}},
  \bibinfo {author} {\bibfnamefont {Y.}~\bibnamefont {Krupko}}, \bibinfo
  {author} {\bibfnamefont {O.}~\bibnamefont {Buisson}}, \bibinfo {author}
  {\bibfnamefont {C.}~\bibnamefont {Naud}}, \bibinfo {author} {\bibfnamefont
  {W.}~\bibnamefont {Hasch-Guichard}}, \bibinfo {author} {\bibfnamefont
  {S.}~\bibnamefont {Florens}}, \bibinfo {author} {\bibfnamefont
  {I.}~\bibnamefont {Snyman}},\ and\ \bibinfo {author} {\bibfnamefont
  {N.}~\bibnamefont {Roch}},\ }\bibfield  {title} {\bibinfo {title} {A tunable
  josephson platform to explore many-body quantum optics in circuit-qed},\
  }\href {https://doi.org/10.1038/s41534-018-0104-0} {\bibfield  {journal}
  {\bibinfo  {journal} {NPJ Quantum Information}\ }\textbf {\bibinfo {volume}
  {5}},\ \bibinfo {pages} {19} (\bibinfo {year} {2019})}\BibitemShut {NoStop}%
\bibitem [{\citenamefont {Roy}\ \emph {et~al.}(2021)\citenamefont {Roy},
  \citenamefont {Schuricht}, \citenamefont {Hauschild}, \citenamefont
  {Pollmann},\ and\ \citenamefont {Saleur}}]{Roy2020b}%
  \BibitemOpen
  \bibfield  {author} {\bibinfo {author} {\bibfnamefont {A.}~\bibnamefont
  {Roy}}, \bibinfo {author} {\bibfnamefont {D.}~\bibnamefont {Schuricht}},
  \bibinfo {author} {\bibfnamefont {J.}~\bibnamefont {Hauschild}}, \bibinfo
  {author} {\bibfnamefont {F.}~\bibnamefont {Pollmann}},\ and\ \bibinfo
  {author} {\bibfnamefont {H.}~\bibnamefont {Saleur}},\ }\bibfield  {title}
  {\bibinfo {title} {{The quantum sine-Gordon model with quantum circuits}},\
  }\href {https://doi.org/10.1016/j.nuclphysb.2021.115445} {\bibfield
  {journal} {\bibinfo  {journal} {Nucl. Phys. B}\ }\textbf {\bibinfo {volume}
  {968}},\ \bibinfo {pages} {115445} (\bibinfo {year} {2021})},\ \Eprint
  {https://arxiv.org/abs/quant-ph/2007.06874} {arXiv:quant-ph/2007.06874}
  \BibitemShut {NoStop}%
\bibitem [{\citenamefont {Roy}\ and\ \citenamefont {Lukyanov}(2023)}]{Roy2023}%
  \BibitemOpen
  \bibfield  {author} {\bibinfo {author} {\bibfnamefont {A.}~\bibnamefont
  {Roy}}\ and\ \bibinfo {author} {\bibfnamefont {S.}~\bibnamefont {Lukyanov}},\
  }\bibfield  {title} {\bibinfo {title} {{Soliton Confinement in a Quantum
  Circuit}},\ }\href@noop {} {\bibfield  {journal} {\bibinfo  {journal}
  {arXiv}\ } (\bibinfo {year} {2023})},\ \Eprint
  {https://arxiv.org/abs/2302.06289} {arXiv:2302.06289 [quant-ph]} \BibitemShut
  {NoStop}%
\bibitem [{\citenamefont {Bradley}\ and\ \citenamefont
  {Doniach}(1984)}]{Bradley1984}%
  \BibitemOpen
  \bibfield  {author} {\bibinfo {author} {\bibfnamefont {R.~M.}\ \bibnamefont
  {Bradley}}\ and\ \bibinfo {author} {\bibfnamefont {S.}~\bibnamefont
  {Doniach}},\ }\bibfield  {title} {\bibinfo {title} {Quantum fluctuations in
  chains of josephson junctions},\ }\href
  {https://doi.org/10.1103/PhysRevB.30.1138} {\bibfield  {journal} {\bibinfo
  {journal} {Phys. Rev. B}\ }\textbf {\bibinfo {volume} {30}},\ \bibinfo
  {pages} {1138} (\bibinfo {year} {1984})}\BibitemShut {NoStop}%
\bibitem [{\citenamefont {Glazman}\ and\ \citenamefont
  {Larkin}(1997)}]{Glazman1997}%
  \BibitemOpen
  \bibfield  {author} {\bibinfo {author} {\bibfnamefont {L.~I.}\ \bibnamefont
  {Glazman}}\ and\ \bibinfo {author} {\bibfnamefont {A.~I.}\ \bibnamefont
  {Larkin}},\ }\bibfield  {title} {\bibinfo {title} {New quantum phase in a
  one-dimensional josephson array},\ }\href
  {https://doi.org/10.1103/PhysRevLett.79.3736} {\bibfield  {journal} {\bibinfo
   {journal} {Phys. Rev. Lett.}\ }\textbf {\bibinfo {volume} {79}},\ \bibinfo
  {pages} {3736} (\bibinfo {year} {1997})}\BibitemShut {NoStop}%
\bibitem [{\citenamefont {Francesco}\ \emph {et~al.}(1997)\citenamefont
  {Francesco}, \citenamefont {Di~Francesco}, \citenamefont {Mathieu},
  \citenamefont {S{\'e}n{\'e}chal},\ and\ \citenamefont
  {Senechal}}]{diFrancesco1997}%
  \BibitemOpen
  \bibfield  {author} {\bibinfo {author} {\bibfnamefont {P.}~\bibnamefont
  {Francesco}}, \bibinfo {author} {\bibfnamefont {P.}~\bibnamefont
  {Di~Francesco}}, \bibinfo {author} {\bibfnamefont {P.}~\bibnamefont
  {Mathieu}}, \bibinfo {author} {\bibfnamefont {D.}~\bibnamefont
  {S{\'e}n{\'e}chal}},\ and\ \bibinfo {author} {\bibfnamefont {D.}~\bibnamefont
  {Senechal}},\ }\href {https://books.google.de/books?id=keUrdME5rhIC} {\emph
  {\bibinfo {title} {Conformal Field Theory}}},\ Graduate Texts in Contemporary
  Physics\ (\bibinfo  {publisher} {Springer},\ \bibinfo {year}
  {1997})\BibitemShut {NoStop}%
\bibitem [{\citenamefont {Belavin}\ \emph {et~al.}(1984)\citenamefont
  {Belavin}, \citenamefont {Polyakov},\ and\ \citenamefont
  {Zamolodchikov}}]{Belavin1984}%
  \BibitemOpen
  \bibfield  {author} {\bibinfo {author} {\bibfnamefont {A.}~\bibnamefont
  {Belavin}}, \bibinfo {author} {\bibfnamefont {A.}~\bibnamefont {Polyakov}},\
  and\ \bibinfo {author} {\bibfnamefont {A.}~\bibnamefont {Zamolodchikov}},\
  }\bibfield  {title} {\bibinfo {title} {Infinite conformal symmetry in
  two-dimensional quantum field theory},\ }\href
  {https://doi.org/https://doi.org/10.1016/0550-3213(84)90052-X} {\bibfield
  {journal} {\bibinfo  {journal} {Nuclear Physics B}\ }\textbf {\bibinfo
  {volume} {241}},\ \bibinfo {pages} {333 } (\bibinfo {year}
  {1984})}\BibitemShut {NoStop}%
\bibitem [{\citenamefont {Zamolodchikov}(1987)}]{Zamolodchikov1987}%
  \BibitemOpen
  \bibfield  {author} {\bibinfo {author} {\bibfnamefont {A.~B.}\ \bibnamefont
  {Zamolodchikov}},\ }\bibfield  {title} {\bibinfo {title} {{Higher Order
  Integrals of Motion in Two-Dimensional Models of the Field Theory with a
  Broken Conformal Symmetry}},\ }\href@noop {} {\bibfield  {journal} {\bibinfo
  {journal} {JETP Lett.}\ }\textbf {\bibinfo {volume} {46}},\ \bibinfo {pages}
  {160} (\bibinfo {year} {1987})},\ \bibinfo {note} {[Pisma Zh. Eksp. Teor.
  Fiz.46,129(1987)]}\BibitemShut {NoStop}%
\bibitem [{\citenamefont {Zamolodchikov}(1989)}]{Zamolodchikov1989}%
  \BibitemOpen
  \bibfield  {author} {\bibinfo {author} {\bibfnamefont {A.}~\bibnamefont
  {Zamolodchikov}},\ }\bibfield  {title} {\bibinfo {title} {Integrable field
  theory from conformal field theory},\ }in\ \href
  {https://doi.org/https://doi.org/10.1016/B978-0-12-385342-4.50022-6} {\emph
  {\bibinfo {booktitle} {Integrable Sys Quantum Field Theory}}},\ \bibinfo
  {editor} {edited by\ \bibinfo {editor} {\bibfnamefont {M.}~\bibnamefont
  {Jimbo}}, \bibinfo {editor} {\bibfnamefont {T.}~\bibnamefont {Miwa}},\ and\
  \bibinfo {editor} {\bibfnamefont {A.}~\bibnamefont {Tsuchiya}}}\ (\bibinfo
  {publisher} {Academic Press},\ \bibinfo {address} {San Diego},\ \bibinfo
  {year} {1989})\ pp.\ \bibinfo {pages} {641 -- 674}\BibitemShut {NoStop}%
\bibitem [{\citenamefont {Teo}\ and\ \citenamefont {Kane}(2014)}]{Teo2014}%
  \BibitemOpen
  \bibfield  {author} {\bibinfo {author} {\bibfnamefont {J.~C.~Y.}\
  \bibnamefont {Teo}}\ and\ \bibinfo {author} {\bibfnamefont {C.~L.}\
  \bibnamefont {Kane}},\ }\bibfield  {title} {\bibinfo {title} {From luttinger
  liquid to non-abelian quantum hall states},\ }\href
  {https://doi.org/10.1103/PhysRevB.89.085101} {\bibfield  {journal} {\bibinfo
  {journal} {Phys. Rev. B}\ }\textbf {\bibinfo {volume} {89}},\ \bibinfo
  {pages} {085101} (\bibinfo {year} {2014})}\BibitemShut {NoStop}%
\bibitem [{\citenamefont {Mong}\ \emph {et~al.}(2014)\citenamefont {Mong},
  \citenamefont {Clarke}, \citenamefont {Alicea}, \citenamefont {Lindner},
  \citenamefont {Fendley}, \citenamefont {Nayak}, \citenamefont {Oreg},
  \citenamefont {Stern}, \citenamefont {Berg}, \citenamefont {Shtengel},\ and\
  \citenamefont {Fisher}}]{Mong2014}%
  \BibitemOpen
  \bibfield  {author} {\bibinfo {author} {\bibfnamefont {R.~S.~K.}\
  \bibnamefont {Mong}}, \bibinfo {author} {\bibfnamefont {D.~J.}\ \bibnamefont
  {Clarke}}, \bibinfo {author} {\bibfnamefont {J.}~\bibnamefont {Alicea}},
  \bibinfo {author} {\bibfnamefont {N.~H.}\ \bibnamefont {Lindner}}, \bibinfo
  {author} {\bibfnamefont {P.}~\bibnamefont {Fendley}}, \bibinfo {author}
  {\bibfnamefont {C.}~\bibnamefont {Nayak}}, \bibinfo {author} {\bibfnamefont
  {Y.}~\bibnamefont {Oreg}}, \bibinfo {author} {\bibfnamefont {A.}~\bibnamefont
  {Stern}}, \bibinfo {author} {\bibfnamefont {E.}~\bibnamefont {Berg}},
  \bibinfo {author} {\bibfnamefont {K.}~\bibnamefont {Shtengel}},\ and\
  \bibinfo {author} {\bibfnamefont {M.~P.~A.}\ \bibnamefont {Fisher}},\
  }\bibfield  {title} {\bibinfo {title} {Universal topological quantum
  computation from a superconductor-abelian quantum hall heterostructure},\
  }\href {https://doi.org/10.1103/PhysRevX.4.011036} {\bibfield  {journal}
  {\bibinfo  {journal} {Phys. Rev. X}\ }\textbf {\bibinfo {volume} {4}},\
  \bibinfo {pages} {011036} (\bibinfo {year} {2014})}\BibitemShut {NoStop}%
\bibitem [{\citenamefont {Kitaev}(2003)}]{Kitaev2003}%
  \BibitemOpen
  \bibfield  {author} {\bibinfo {author} {\bibfnamefont {A.}~\bibnamefont
  {Kitaev}},\ }\bibfield  {title} {\bibinfo {title} {{Fault-tolerant quantum
  computation by anyons}},\ }\href
  {https://doi.org/10.1016/S0003-4916(02)00018-0} {\bibfield  {journal}
  {\bibinfo  {journal} {Ann. Phys. (NY)}\ }\textbf {\bibinfo {volume} {303}},\
  \bibinfo {pages} {2} (\bibinfo {year} {2003})}\BibitemShut {NoStop}%
\bibitem [{\citenamefont {Andrews}\ \emph {et~al.}(1984)\citenamefont
  {Andrews}, \citenamefont {Baxter},\ and\ \citenamefont
  {Forrester}}]{Andrews:1984af}%
  \BibitemOpen
  \bibfield  {author} {\bibinfo {author} {\bibfnamefont {G.~E.}\ \bibnamefont
  {Andrews}}, \bibinfo {author} {\bibfnamefont {R.~J.}\ \bibnamefont
  {Baxter}},\ and\ \bibinfo {author} {\bibfnamefont {P.~J.}\ \bibnamefont
  {Forrester}},\ }\bibfield  {title} {\bibinfo {title} {{Eight vertex SOS model
  and generalized Rogers-Ramanujan type identities}},\ }\href
  {https://doi.org/10.1007/BF01014383} {\bibfield  {journal} {\bibinfo
  {journal} {J. Statist. Phys.}\ }\textbf {\bibinfo {volume} {35}},\ \bibinfo
  {pages} {193} (\bibinfo {year} {1984})}\BibitemShut {NoStop}%
\bibitem [{\citenamefont {Bazhanov}\ and\ \citenamefont
  {Reshetikhin}(1990)}]{Bazhanov_1990}%
  \BibitemOpen
  \bibfield  {author} {\bibinfo {author} {\bibfnamefont {V.~V.}\ \bibnamefont
  {Bazhanov}}\ and\ \bibinfo {author} {\bibfnamefont {N.}~\bibnamefont
  {Reshetikhin}},\ }\bibfield  {title} {\bibinfo {title} {Restricted
  solid-on-solid models connected with simply laced algebras and conformal
  field theory},\ }\href {https://doi.org/10.1088/0305-4470/23/9/012}
  {\bibfield  {journal} {\bibinfo  {journal} {Journal of Physics A:
  Mathematical and General}\ }\textbf {\bibinfo {volume} {23}},\ \bibinfo
  {pages} {1477} (\bibinfo {year} {1990})}\BibitemShut {NoStop}%
\bibitem [{\citenamefont {Blume}(1966)}]{Blume1966}%
  \BibitemOpen
  \bibfield  {author} {\bibinfo {author} {\bibfnamefont {M.}~\bibnamefont
  {Blume}},\ }\bibfield  {title} {\bibinfo {title} {Theory of the first-order
  magnetic phase change in u${\mathrm{o}}_{2}$},\ }\href
  {https://doi.org/10.1103/PhysRev.141.517} {\bibfield  {journal} {\bibinfo
  {journal} {Phys. Rev.}\ }\textbf {\bibinfo {volume} {141}},\ \bibinfo {pages}
  {517} (\bibinfo {year} {1966})}\BibitemShut {NoStop}%
\bibitem [{\citenamefont {Capel}(1966)}]{Capel1966}%
  \BibitemOpen
  \bibfield  {author} {\bibinfo {author} {\bibfnamefont {H.}~\bibnamefont
  {Capel}},\ }\bibfield  {title} {\bibinfo {title} {On the possibility of
  first-order phase transitions in ising systems of triplet ions with
  zero-field splitting},\ }\href
  {https://doi.org/https://doi.org/10.1016/0031-8914(66)90027-9} {\bibfield
  {journal} {\bibinfo  {journal} {Physica}\ }\textbf {\bibinfo {volume} {32}},\
  \bibinfo {pages} {966} (\bibinfo {year} {1966})}\BibitemShut {NoStop}%
\bibitem [{\citenamefont {Rahmani}\ \emph {et~al.}(2015)\citenamefont
  {Rahmani}, \citenamefont {Zhu}, \citenamefont {Franz},\ and\ \citenamefont
  {Affleck}}]{Rahmani2015}%
  \BibitemOpen
  \bibfield  {author} {\bibinfo {author} {\bibfnamefont {A.}~\bibnamefont
  {Rahmani}}, \bibinfo {author} {\bibfnamefont {X.}~\bibnamefont {Zhu}},
  \bibinfo {author} {\bibfnamefont {M.}~\bibnamefont {Franz}},\ and\ \bibinfo
  {author} {\bibfnamefont {I.}~\bibnamefont {Affleck}},\ }\bibfield  {title}
  {\bibinfo {title} {Phase diagram of the interacting majorana chain model},\
  }\href {https://doi.org/10.1103/PhysRevB.92.235123} {\bibfield  {journal}
  {\bibinfo  {journal} {Phys. Rev. B}\ }\textbf {\bibinfo {volume} {92}},\
  \bibinfo {pages} {235123} (\bibinfo {year} {2015})}\BibitemShut {NoStop}%
\bibitem [{\citenamefont {O'Brien}\ and\ \citenamefont
  {Fendley}(2018)}]{Brien2018}%
  \BibitemOpen
  \bibfield  {author} {\bibinfo {author} {\bibfnamefont {E.}~\bibnamefont
  {O'Brien}}\ and\ \bibinfo {author} {\bibfnamefont {P.}~\bibnamefont
  {Fendley}},\ }\bibfield  {title} {\bibinfo {title} {Lattice supersymmetry and
  order-disorder coexistence in the tricritical ising model},\ }\href
  {https://doi.org/10.1103/PhysRevLett.120.206403} {\bibfield  {journal}
  {\bibinfo  {journal} {Phys. Rev. Lett.}\ }\textbf {\bibinfo {volume} {120}},\
  \bibinfo {pages} {206403} (\bibinfo {year} {2018})}\BibitemShut {NoStop}%
\bibitem [{DMR()}]{DMRG_TeNPy}%
  \BibitemOpen
  \href@noop {} {}\bibinfo {note} {The DMRG computations of this work were
  performed using the TeNPy package~\cite{Hauschild2018}.}\BibitemShut {Stop}%
\bibitem [{\citenamefont {Lukyanov}\ and\ \citenamefont
  {Terras}(2003)}]{Lukyanov2003}%
  \BibitemOpen
  \bibfield  {author} {\bibinfo {author} {\bibfnamefont {S.}~\bibnamefont
  {Lukyanov}}\ and\ \bibinfo {author} {\bibfnamefont {V.}~\bibnamefont
  {Terras}},\ }\bibfield  {title} {\bibinfo {title} {Long-distance asymptotics
  of spin–spin correlation functions for the xxz spin chain},\ }\href
  {https://doi.org/https://doi.org/10.1016/S0550-3213(02)01141-0} {\bibfield
  {journal} {\bibinfo  {journal} {Nuclear Physics B}\ }\textbf {\bibinfo
  {volume} {654}},\ \bibinfo {pages} {323 } (\bibinfo {year}
  {2003})}\BibitemShut {NoStop}%
\bibitem [{\citenamefont {Baxter}(2013)}]{Baxter2013}%
  \BibitemOpen
  \bibfield  {author} {\bibinfo {author} {\bibfnamefont {R.}~\bibnamefont
  {Baxter}},\ }\href {https://books.google.de/books?id=eQzCAgAAQBAJ} {\emph
  {\bibinfo {title} {Exactly Solved Models in Statistical Mechanics}}},\ Dover
  Books on Physics\ (\bibinfo  {publisher} {Dover Publications},\ \bibinfo
  {year} {2013})\BibitemShut {NoStop}%
\bibitem [{\citenamefont {Zamolodchikov}(1986)}]{Zamolodchikov:1986db}%
  \BibitemOpen
  \bibfield  {author} {\bibinfo {author} {\bibfnamefont {A.~B.}\ \bibnamefont
  {Zamolodchikov}},\ }\bibfield  {title} {\bibinfo {title} {{Conformal Symmetry
  and Multicritical Points in Two-Dimensional Quantum Field Theory. (In
  Russian)}},\ }\href@noop {} {\bibfield  {journal} {\bibinfo  {journal} {Sov.
  J. Nucl. Phys.}\ }\textbf {\bibinfo {volume} {44}},\ \bibinfo {pages} {529}
  (\bibinfo {year} {1986})}\BibitemShut {NoStop}%
\bibitem [{\citenamefont {Roy}\ \emph {et~al.}(2020)\citenamefont {Roy},
  \citenamefont {Pollmann},\ and\ \citenamefont {Saleur}}]{Roy2020a}%
  \BibitemOpen
  \bibfield  {author} {\bibinfo {author} {\bibfnamefont {A.}~\bibnamefont
  {Roy}}, \bibinfo {author} {\bibfnamefont {F.}~\bibnamefont {Pollmann}},\ and\
  \bibinfo {author} {\bibfnamefont {H.}~\bibnamefont {Saleur}},\ }\bibfield
  {title} {\bibinfo {title} {{Entanglement Hamiltonian of the 1+1-dimensional
  free, compactified boson conformal field theory}},\ }\href
  {https://doi.org/10.1088/1742-5468/aba498} {\bibfield  {journal} {\bibinfo
  {journal} {J. Stat. Mech.}\ }\textbf {\bibinfo {volume} {2008}},\ \bibinfo
  {pages} {083104} (\bibinfo {year} {2020})},\ \Eprint
  {https://arxiv.org/abs/cond-mat/2004.14370} {arXiv:cond-mat/2004.14370}
  \BibitemShut {NoStop}%
\bibitem [{\citenamefont {Delfino}\ and\ \citenamefont
  {Mussardo}(1998)}]{Delfino1998}%
  \BibitemOpen
  \bibfield  {author} {\bibinfo {author} {\bibfnamefont {G.}~\bibnamefont
  {Delfino}}\ and\ \bibinfo {author} {\bibfnamefont {G.}~\bibnamefont
  {Mussardo}},\ }\bibfield  {title} {\bibinfo {title} {Non-integrable aspects
  of the multi-frequency sine-gordon model},\ }\href
  {https://doi.org/10.1016/s0550-3213(98)00063-7} {\bibfield  {journal}
  {\bibinfo  {journal} {Nuclear Physics B}\ }\textbf {\bibinfo {volume}
  {516}},\ \bibinfo {pages} {675–703} (\bibinfo {year} {1998})}\BibitemShut
  {NoStop}%
\bibitem [{\citenamefont {Mussardo}\ \emph {et~al.}(2004)\citenamefont
  {Mussardo}, \citenamefont {Riva},\ and\ \citenamefont
  {Sotkov}}]{Mussardo2004}%
  \BibitemOpen
  \bibfield  {author} {\bibinfo {author} {\bibfnamefont {G.}~\bibnamefont
  {Mussardo}}, \bibinfo {author} {\bibfnamefont {V.}~\bibnamefont {Riva}},\
  and\ \bibinfo {author} {\bibfnamefont {G.}~\bibnamefont {Sotkov}},\
  }\bibfield  {title} {\bibinfo {title} {{Semiclassical particle spectrum of
  double sine-Gordon model}},\ }\href
  {https://doi.org/10.1016/j.nuclphysb.2004.04.003} {\bibfield  {journal}
  {\bibinfo  {journal} {Nucl. Phys. B}\ }\textbf {\bibinfo {volume} {687}},\
  \bibinfo {pages} {189} (\bibinfo {year} {2004})},\ \Eprint
  {https://arxiv.org/abs/hep-th/0402179} {arXiv:hep-th/0402179} \BibitemShut
  {NoStop}%
\bibitem [{\citenamefont {Bajnok}\ \emph {et~al.}(2001)\citenamefont {Bajnok},
  \citenamefont {Palla}, \citenamefont {Takacs},\ and\ \citenamefont
  {Wagner}}]{Bajnok2000}%
  \BibitemOpen
  \bibfield  {author} {\bibinfo {author} {\bibfnamefont {Z.}~\bibnamefont
  {Bajnok}}, \bibinfo {author} {\bibfnamefont {L.}~\bibnamefont {Palla}},
  \bibinfo {author} {\bibfnamefont {G.}~\bibnamefont {Takacs}},\ and\ \bibinfo
  {author} {\bibfnamefont {F.}~\bibnamefont {Wagner}},\ }\bibfield  {title}
  {\bibinfo {title} {{Nonperturbative study of the two frequency sine-Gordon
  model}},\ }\href {https://doi.org/10.1016/S0550-3213(01)00067-0} {\bibfield
  {journal} {\bibinfo  {journal} {Nucl. Phys. B}\ }\textbf {\bibinfo {volume}
  {601}},\ \bibinfo {pages} {503} (\bibinfo {year} {2001})},\ \Eprint
  {https://arxiv.org/abs/hep-th/0008066} {arXiv:hep-th/0008066} \BibitemShut
  {NoStop}%
\bibitem [{\citenamefont {Dou\ifmmode~\mbox{\c{c}}\else \c{c}\fi{}ot}\ and\
  \citenamefont {Vidal}(2002)}]{Doucot2002}%
  \BibitemOpen
  \bibfield  {author} {\bibinfo {author} {\bibfnamefont {B.}~\bibnamefont
  {Dou\ifmmode~\mbox{\c{c}}\else \c{c}\fi{}ot}}\ and\ \bibinfo {author}
  {\bibfnamefont {J.}~\bibnamefont {Vidal}},\ }\bibfield  {title} {\bibinfo
  {title} {Pairing of cooper pairs in a fully frustrated josephson-junction
  chain},\ }\href {https://doi.org/10.1103/PhysRevLett.88.227005} {\bibfield
  {journal} {\bibinfo  {journal} {Phys. Rev. Lett.}\ }\textbf {\bibinfo
  {volume} {88}},\ \bibinfo {pages} {227005} (\bibinfo {year}
  {2002})}\BibitemShut {NoStop}%
\bibitem [{\citenamefont {Ioffe}\ and\ \citenamefont
  {Feigel'man}(2002)}]{Ioffe2002}%
  \BibitemOpen
  \bibfield  {author} {\bibinfo {author} {\bibfnamefont {L.~B.}\ \bibnamefont
  {Ioffe}}\ and\ \bibinfo {author} {\bibfnamefont {M.~V.}\ \bibnamefont
  {Feigel'man}},\ }\bibfield  {title} {\bibinfo {title} {Possible realization
  of an ideal quantum computer in josephson junction array},\ }\href
  {https://doi.org/10.1103/PhysRevB.66.224503} {\bibfield  {journal} {\bibinfo
  {journal} {Phys. Rev. B}\ }\textbf {\bibinfo {volume} {66}},\ \bibinfo
  {pages} {224503} (\bibinfo {year} {2002})}\BibitemShut {NoStop}%
\bibitem [{\citenamefont {Kitaev}(2006)}]{Kitaev2006c}%
  \BibitemOpen
  \bibfield  {author} {\bibinfo {author} {\bibfnamefont {A.}~\bibnamefont
  {Kitaev}},\ }\href {https://doi.org/10.48550/ARXIV.COND-MAT/0609441}
  {\bibinfo {title} {Protected qubit based on a superconducting current
  mirror}} (\bibinfo {year} {2006}),\ \Eprint
  {https://arxiv.org/abs/cond-mat/0609441} {arXiv:cond-mat/0609441}
  \BibitemShut {NoStop}%
\bibitem [{\citenamefont {Brooks}\ \emph {et~al.}(2013)\citenamefont {Brooks},
  \citenamefont {Kitaev},\ and\ \citenamefont {Preskill}}]{Brooks2013}%
  \BibitemOpen
  \bibfield  {author} {\bibinfo {author} {\bibfnamefont {P.}~\bibnamefont
  {Brooks}}, \bibinfo {author} {\bibfnamefont {A.}~\bibnamefont {Kitaev}},\
  and\ \bibinfo {author} {\bibfnamefont {J.}~\bibnamefont {Preskill}},\
  }\bibfield  {title} {\bibinfo {title} {Protected gates for superconducting
  qubits},\ }\href {https://doi.org/10.1103/PhysRevA.87.052306} {\bibfield
  {journal} {\bibinfo  {journal} {Phys. Rev. A}\ }\textbf {\bibinfo {volume}
  {87}},\ \bibinfo {pages} {052306} (\bibinfo {year} {2013})}\BibitemShut
  {NoStop}%
\bibitem [{\citenamefont {Gladchenko}\ \emph {et~al.}(2008)\citenamefont
  {Gladchenko}, \citenamefont {Olaya}, \citenamefont {Dupont-Ferrier},
  \citenamefont {Dou{\c{c}}ot}, \citenamefont {Ioffe},\ and\ \citenamefont
  {Gershenson}}]{Gladchenko2008}%
  \BibitemOpen
  \bibfield  {author} {\bibinfo {author} {\bibfnamefont {S.}~\bibnamefont
  {Gladchenko}}, \bibinfo {author} {\bibfnamefont {D.}~\bibnamefont {Olaya}},
  \bibinfo {author} {\bibfnamefont {E.}~\bibnamefont {Dupont-Ferrier}},
  \bibinfo {author} {\bibfnamefont {B.}~\bibnamefont {Dou{\c{c}}ot}}, \bibinfo
  {author} {\bibfnamefont {L.~B.}\ \bibnamefont {Ioffe}},\ and\ \bibinfo
  {author} {\bibfnamefont {M.~E.}\ \bibnamefont {Gershenson}},\ }\bibfield
  {title} {\bibinfo {title} {Superconducting nanocircuits for topologically
  protected qubits},\ }\href {https://doi.org/10.1038/nphys1151} {\bibfield
  {journal} {\bibinfo  {journal} {Nature Physics}\ }\textbf {\bibinfo {volume}
  {5}},\ \bibinfo {pages} {48} (\bibinfo {year} {2008})}\BibitemShut {NoStop}%
\bibitem [{\citenamefont {{Smith}}\ \emph {et~al.}(2020)\citenamefont
  {{Smith}}, \citenamefont {{Kou}}, \citenamefont {{Xiao}}, \citenamefont
  {{Vool}},\ and\ \citenamefont {{Devoret}}}]{Smith2020}%
  \BibitemOpen
  \bibfield  {author} {\bibinfo {author} {\bibfnamefont {W.~C.}\ \bibnamefont
  {{Smith}}}, \bibinfo {author} {\bibfnamefont {A.}~\bibnamefont {{Kou}}},
  \bibinfo {author} {\bibfnamefont {X.}~\bibnamefont {{Xiao}}}, \bibinfo
  {author} {\bibfnamefont {U.}~\bibnamefont {{Vool}}},\ and\ \bibinfo {author}
  {\bibfnamefont {M.~H.}\ \bibnamefont {{Devoret}}},\ }\bibfield  {title}
  {\bibinfo {title} {{Superconducting circuit protected by two-Cooper-pair
  tunneling}},\ }\href {https://doi.org/10.1038/s41534-019-0231-2} {\bibfield
  {journal} {\bibinfo  {journal} {NPJ Quantum Information}\ }\textbf {\bibinfo
  {volume} {6}},\ \bibinfo {eid} {8} (\bibinfo {year} {2020})},\ \Eprint
  {https://arxiv.org/abs/1905.01206} {arXiv:1905.01206 [quant-ph]} \BibitemShut
  {NoStop}%
\bibitem [{\citenamefont {Gyenis}\ \emph {et~al.}(2021)\citenamefont {Gyenis},
  \citenamefont {Mundada}, \citenamefont {Di~Paolo}, \citenamefont {Hazard},
  \citenamefont {You}, \citenamefont {Schuster}, \citenamefont {Koch},
  \citenamefont {Blais},\ and\ \citenamefont {Houck}}]{Gyenis2021}%
  \BibitemOpen
  \bibfield  {author} {\bibinfo {author} {\bibfnamefont {A.}~\bibnamefont
  {Gyenis}}, \bibinfo {author} {\bibfnamefont {P.~S.}\ \bibnamefont {Mundada}},
  \bibinfo {author} {\bibfnamefont {A.}~\bibnamefont {Di~Paolo}}, \bibinfo
  {author} {\bibfnamefont {T.~M.}\ \bibnamefont {Hazard}}, \bibinfo {author}
  {\bibfnamefont {X.}~\bibnamefont {You}}, \bibinfo {author} {\bibfnamefont
  {D.~I.}\ \bibnamefont {Schuster}}, \bibinfo {author} {\bibfnamefont
  {J.}~\bibnamefont {Koch}}, \bibinfo {author} {\bibfnamefont {A.}~\bibnamefont
  {Blais}},\ and\ \bibinfo {author} {\bibfnamefont {A.~A.}\ \bibnamefont
  {Houck}},\ }\bibfield  {title} {\bibinfo {title} {Experimental realization of
  a protected superconducting circuit derived from the $0$--$\ensuremath{\pi}$
  qubit},\ }\href {https://doi.org/10.1103/PRXQuantum.2.010339} {\bibfield
  {journal} {\bibinfo  {journal} {PRX Quantum}\ }\textbf {\bibinfo {volume}
  {2}},\ \bibinfo {pages} {010339} (\bibinfo {year} {2021})}\BibitemShut
  {NoStop}%
\bibitem [{Note1()}]{Note1}%
  \BibitemOpen
  \bibinfo {note} {Note that the effective capacitance~$C_g$ is the sum of the
  two capacitances of the~$\protect \qopname \relax o{cos}\phi $ and
  the~$\protect \qopname \relax o{cos}2\phi $ Josephson junctions.}\BibitemShut
  {Stop}%
\bibitem [{\citenamefont {Lukyanov}\ and\ \citenamefont
  {Zamolodchikov}(2001)}]{Lukyanov2001}%
  \BibitemOpen
  \bibfield  {author} {\bibinfo {author} {\bibfnamefont {S.}~\bibnamefont
  {Lukyanov}}\ and\ \bibinfo {author} {\bibfnamefont {A.}~\bibnamefont
  {Zamolodchikov}},\ }\bibfield  {title} {\bibinfo {title} {Form factors of
  soliton-creating operators in the sine-gordon model},\ }\href
  {https://doi.org/10.1016/s0550-3213(01)00262-0} {\bibfield  {journal}
  {\bibinfo  {journal} {Nuclear Physics B}\ }\textbf {\bibinfo {volume}
  {607}},\ \bibinfo {pages} {437} (\bibinfo {year} {2001})}\BibitemShut
  {NoStop}%
\bibitem [{\citenamefont {Calabrese}\ and\ \citenamefont
  {Cardy}(2004)}]{Calabrese2004}%
  \BibitemOpen
  \bibfield  {author} {\bibinfo {author} {\bibfnamefont {P.}~\bibnamefont
  {Calabrese}}\ and\ \bibinfo {author} {\bibfnamefont {J.}~\bibnamefont
  {Cardy}},\ }\bibfield  {title} {\bibinfo {title} {Entanglement entropy and
  quantum field theory},\ }\href
  {https://doi.org/10.1088/1742-5468/2004/06/p06002} {\bibfield  {journal}
  {\bibinfo  {journal} {J. Stat. Mech: Theory and Experiment}\ }\textbf
  {\bibinfo {volume} {2004}},\ \bibinfo {pages} {P06002} (\bibinfo {year}
  {2004})}\BibitemShut {NoStop}%
\bibitem [{\citenamefont {Pollmann}\ \emph {et~al.}(2009)\citenamefont
  {Pollmann}, \citenamefont {Mukerjee}, \citenamefont {Turner},\ and\
  \citenamefont {Moore}}]{Pollmann2009}%
  \BibitemOpen
  \bibfield  {author} {\bibinfo {author} {\bibfnamefont {F.}~\bibnamefont
  {Pollmann}}, \bibinfo {author} {\bibfnamefont {S.}~\bibnamefont {Mukerjee}},
  \bibinfo {author} {\bibfnamefont {A.~M.}\ \bibnamefont {Turner}},\ and\
  \bibinfo {author} {\bibfnamefont {J.~E.}\ \bibnamefont {Moore}},\ }\bibfield
  {title} {\bibinfo {title} {Theory of finite-entanglement scaling at
  one-dimensional quantum critical points},\ }\href
  {https://doi.org/10.1103/PhysRevLett.102.255701} {\bibfield  {journal}
  {\bibinfo  {journal} {Phys. Rev. Lett.}\ }\textbf {\bibinfo {volume} {102}},\
  \bibinfo {pages} {255701} (\bibinfo {year} {2009})}\BibitemShut {NoStop}%
\bibitem [{\citenamefont {Ghoshal}\ and\ \citenamefont
  {Zamolodchikov}(1994)}]{Ghoshal1994}%
  \BibitemOpen
  \bibfield  {author} {\bibinfo {author} {\bibfnamefont {S.}~\bibnamefont
  {Ghoshal}}\ and\ \bibinfo {author} {\bibfnamefont {A.}~\bibnamefont
  {Zamolodchikov}},\ }\bibfield  {title} {\bibinfo {title} {Boundary s matrix
  and boundary state in two-dimensional integrable quantum field theory},\
  }\href {https://doi.org/10.1142/S0217751X94001552} {\bibfield  {journal}
  {\bibinfo  {journal} {International Journal of Modern Physics A}\ }\textbf
  {\bibinfo {volume} {09}},\ \bibinfo {pages} {3841} (\bibinfo {year}
  {1994})},\ \Eprint
  {https://arxiv.org/abs/https://doi.org/10.1142/S0217751X94001552}
  {https://doi.org/10.1142/S0217751X94001552} \BibitemShut {NoStop}%
\bibitem [{\citenamefont {Fendley}\ \emph {et~al.}(1994)\citenamefont
  {Fendley}, \citenamefont {Saleur},\ and\ \citenamefont
  {Warner}}]{Fendley1994}%
  \BibitemOpen
  \bibfield  {author} {\bibinfo {author} {\bibfnamefont {P.}~\bibnamefont
  {Fendley}}, \bibinfo {author} {\bibfnamefont {H.}~\bibnamefont {Saleur}},\
  and\ \bibinfo {author} {\bibfnamefont {N.}~\bibnamefont {Warner}},\
  }\bibfield  {title} {\bibinfo {title} {Exact solution of a massless scalar
  field with a relevant boundary interaction},\ }\href
  {https://doi.org/https://doi.org/10.1016/0550-3213(94)90160-0} {\bibfield
  {journal} {\bibinfo  {journal} {Nuclear Physics B}\ }\textbf {\bibinfo
  {volume} {430}},\ \bibinfo {pages} {577 } (\bibinfo {year}
  {1994})}\BibitemShut {NoStop}%
\bibitem [{\citenamefont {Affleck}\ and\ \citenamefont
  {Ludwig}(1991)}]{Affleck1991}%
  \BibitemOpen
  \bibfield  {author} {\bibinfo {author} {\bibfnamefont {I.}~\bibnamefont
  {Affleck}}\ and\ \bibinfo {author} {\bibfnamefont {A.~W.~W.}\ \bibnamefont
  {Ludwig}},\ }\bibfield  {title} {\bibinfo {title} {Universal noninteger
  ``ground-state degeneracy'' in critical quantum systems},\ }\href
  {https://doi.org/10.1103/PhysRevLett.67.161} {\bibfield  {journal} {\bibinfo
  {journal} {Phys. Rev. Lett.}\ }\textbf {\bibinfo {volume} {67}},\ \bibinfo
  {pages} {161} (\bibinfo {year} {1991})}\BibitemShut {NoStop}%
\bibitem [{\citenamefont {Saleur}(1998)}]{Saleur1998}%
  \BibitemOpen
  \bibfield  {author} {\bibinfo {author} {\bibfnamefont {H.}~\bibnamefont
  {Saleur}},\ }\bibfield  {title} {\bibinfo {title} {{Lectures on
  nonperturbative field theory and quantum impurity problems}},\ }\href@noop {}
  {\  (\bibinfo {year} {1998})},\ \Eprint
  {https://arxiv.org/abs/cond-mat/9812110} {arXiv:cond-mat/9812110}
  \BibitemShut {NoStop}%
\bibitem [{\citenamefont {Lassig}\ \emph {et~al.}(1991)\citenamefont {Lassig},
  \citenamefont {Mussardo},\ and\ \citenamefont {Cardy}}]{Lassig:1990xy}%
  \BibitemOpen
  \bibfield  {author} {\bibinfo {author} {\bibfnamefont {M.}~\bibnamefont
  {Lassig}}, \bibinfo {author} {\bibfnamefont {G.}~\bibnamefont {Mussardo}},\
  and\ \bibinfo {author} {\bibfnamefont {J.~L.}\ \bibnamefont {Cardy}},\
  }\bibfield  {title} {\bibinfo {title} {{The scaling region of the tricritical
  Ising model in two-dimensions}},\ }\href
  {https://doi.org/10.1016/0550-3213(91)90206-D} {\bibfield  {journal}
  {\bibinfo  {journal} {Nucl. Phys. B}\ }\textbf {\bibinfo {volume} {348}},\
  \bibinfo {pages} {591} (\bibinfo {year} {1991})}\BibitemShut {NoStop}%
\bibitem [{\citenamefont {Lencs\'es}\ \emph {et~al.}(2023)\citenamefont
  {Lencs\'es}, \citenamefont {Mussardo},\ and\ \citenamefont
  {Tak\'acs}}]{Lencses:2023hlq}%
  \BibitemOpen
  \bibfield  {author} {\bibinfo {author} {\bibfnamefont {M.}~\bibnamefont
  {Lencs\'es}}, \bibinfo {author} {\bibfnamefont {G.}~\bibnamefont
  {Mussardo}},\ and\ \bibinfo {author} {\bibfnamefont {G.}~\bibnamefont
  {Tak\'acs}},\ }\bibfield  {title} {\bibinfo {title} {{Quantum Integrability
  vs Experiments: Correlation Functions and Dynamical Structure Factors}},\
  }\href@noop {} {\  (\bibinfo {year} {2023})},\ \Eprint
  {https://arxiv.org/abs/2303.16556} {arXiv:2303.16556 [cond-mat.stat-mech]}
  \BibitemShut {NoStop}%
\bibitem [{\citenamefont {Caldeira}\ and\ \citenamefont
  {Leggett}(1983)}]{Caldeira1983}%
  \BibitemOpen
  \bibfield  {author} {\bibinfo {author} {\bibfnamefont {A.}~\bibnamefont
  {Caldeira}}\ and\ \bibinfo {author} {\bibfnamefont {A.}~\bibnamefont
  {Leggett}},\ }\bibfield  {title} {\bibinfo {title} {Path integral approach to
  quantum brownian motion},\ }\href
  {https://doi.org/https://doi.org/10.1016/0378-4371(83)90013-4} {\bibfield
  {journal} {\bibinfo  {journal} {Physica A: Statistical Mechanics and its
  Applications}\ }\textbf {\bibinfo {volume} {121}},\ \bibinfo {pages} {587}
  (\bibinfo {year} {1983})}\BibitemShut {NoStop}%
\bibitem [{\citenamefont {Schmid}(1983)}]{Schmid1983}%
  \BibitemOpen
  \bibfield  {author} {\bibinfo {author} {\bibfnamefont {A.}~\bibnamefont
  {Schmid}},\ }\bibfield  {title} {\bibinfo {title} {Diffusion and localization
  in a dissipative quantum system},\ }\href
  {https://doi.org/10.1103/PhysRevLett.51.1506} {\bibfield  {journal} {\bibinfo
   {journal} {Phys. Rev. Lett.}\ }\textbf {\bibinfo {volume} {51}},\ \bibinfo
  {pages} {1506} (\bibinfo {year} {1983})}\BibitemShut {NoStop}%
\bibitem [{\citenamefont {Houzet}\ and\ \citenamefont
  {Glazman}(2020)}]{Houzet2020}%
  \BibitemOpen
  \bibfield  {author} {\bibinfo {author} {\bibfnamefont {M.}~\bibnamefont
  {Houzet}}\ and\ \bibinfo {author} {\bibfnamefont {L.~I.}\ \bibnamefont
  {Glazman}},\ }\bibfield  {title} {\bibinfo {title} {Critical fluorescence of
  a transmon at the schmid transition},\ }\href
  {https://doi.org/10.1103/PhysRevLett.125.267701} {\bibfield  {journal}
  {\bibinfo  {journal} {Phys. Rev. Lett.}\ }\textbf {\bibinfo {volume} {125}},\
  \bibinfo {pages} {267701} (\bibinfo {year} {2020})}\BibitemShut {NoStop}%
\bibitem [{\citenamefont {Petkova}\ and\ \citenamefont
  {Zuber}(2001)}]{Petkova:2000ip}%
  \BibitemOpen
  \bibfield  {author} {\bibinfo {author} {\bibfnamefont {V.~B.}\ \bibnamefont
  {Petkova}}\ and\ \bibinfo {author} {\bibfnamefont {J.~B.}\ \bibnamefont
  {Zuber}},\ }\bibfield  {title} {\bibinfo {title} {{Generalized twisted
  partition functions}},\ }\href
  {https://doi.org/10.1016/S0370-2693(01)00276-3} {\bibfield  {journal}
  {\bibinfo  {journal} {Phys. Lett. B}\ }\textbf {\bibinfo {volume} {504}},\
  \bibinfo {pages} {157} (\bibinfo {year} {2001})},\ \Eprint
  {https://arxiv.org/abs/hep-th/0011021} {arXiv:hep-th/0011021} \BibitemShut
  {NoStop}%
\bibitem [{\citenamefont {Bachas}\ \emph {et~al.}(2002)\citenamefont {Bachas},
  \citenamefont {de~Boer}, \citenamefont {Dijkgraaf},\ and\ \citenamefont
  {Ooguri}}]{Bachas2001}%
  \BibitemOpen
  \bibfield  {author} {\bibinfo {author} {\bibfnamefont {C.}~\bibnamefont
  {Bachas}}, \bibinfo {author} {\bibfnamefont {J.}~\bibnamefont {de~Boer}},
  \bibinfo {author} {\bibfnamefont {R.}~\bibnamefont {Dijkgraaf}},\ and\
  \bibinfo {author} {\bibfnamefont {H.}~\bibnamefont {Ooguri}},\ }\bibfield
  {title} {\bibinfo {title} {{Permeable conformal walls and holography}},\
  }\href {https://doi.org/10.1088/1126-6708/2002/06/027} {\bibfield  {journal}
  {\bibinfo  {journal} {JHEP}\ }\textbf {\bibinfo {volume} {06}},\ \bibinfo
  {pages} {027}},\ \Eprint {https://arxiv.org/abs/hep-th/0111210}
  {arXiv:hep-th/0111210} \BibitemShut {NoStop}%
\bibitem [{\citenamefont {Fr\"ohlich}\ \emph {et~al.}(2004)\citenamefont
  {Fr\"ohlich}, \citenamefont {Fuchs}, \citenamefont {Runkel},\ and\
  \citenamefont {Schweigert}}]{Frohlich2004}%
  \BibitemOpen
  \bibfield  {author} {\bibinfo {author} {\bibfnamefont {J.}~\bibnamefont
  {Fr\"ohlich}}, \bibinfo {author} {\bibfnamefont {J.}~\bibnamefont {Fuchs}},
  \bibinfo {author} {\bibfnamefont {I.}~\bibnamefont {Runkel}},\ and\ \bibinfo
  {author} {\bibfnamefont {C.}~\bibnamefont {Schweigert}},\ }\bibfield  {title}
  {\bibinfo {title} {Kramers-wannier duality from conformal defects},\ }\href
  {https://doi.org/10.1103/PhysRevLett.93.070601} {\bibfield  {journal}
  {\bibinfo  {journal} {Phys. Rev. Lett.}\ }\textbf {\bibinfo {volume} {93}},\
  \bibinfo {pages} {070601} (\bibinfo {year} {2004})}\BibitemShut {NoStop}%
\bibitem [{\citenamefont {Frohlich}\ \emph {et~al.}(2007)\citenamefont
  {Frohlich}, \citenamefont {Fuchs}, \citenamefont {Runkel},\ and\
  \citenamefont {Schweigert}}]{Frohlich2006}%
  \BibitemOpen
  \bibfield  {author} {\bibinfo {author} {\bibfnamefont {J.}~\bibnamefont
  {Frohlich}}, \bibinfo {author} {\bibfnamefont {J.}~\bibnamefont {Fuchs}},
  \bibinfo {author} {\bibfnamefont {I.}~\bibnamefont {Runkel}},\ and\ \bibinfo
  {author} {\bibfnamefont {C.}~\bibnamefont {Schweigert}},\ }\bibfield  {title}
  {\bibinfo {title} {{Duality and defects in rational conformal field
  theory}},\ }\href {https://doi.org/10.1016/j.nuclphysb.2006.11.017}
  {\bibfield  {journal} {\bibinfo  {journal} {Nucl. Phys. B}\ }\textbf
  {\bibinfo {volume} {763}},\ \bibinfo {pages} {354} (\bibinfo {year}
  {2007})},\ \Eprint {https://arxiv.org/abs/hep-th/0607247}
  {arXiv:hep-th/0607247} \BibitemShut {NoStop}%
\bibitem [{\citenamefont {Grimm}(2002)}]{Grimm2001}%
  \BibitemOpen
  \bibfield  {author} {\bibinfo {author} {\bibfnamefont {U.}~\bibnamefont
  {Grimm}},\ }\bibfield  {title} {\bibinfo {title} {{Spectrum of a duality
  twisted Ising quantum chain}},\ }\href
  {https://doi.org/10.1088/0305-4470/35/3/101} {\bibfield  {journal} {\bibinfo
  {journal} {J. Phys. A}\ }\textbf {\bibinfo {volume} {35}},\ \bibinfo {pages}
  {L25} (\bibinfo {year} {2002})},\ \Eprint
  {https://arxiv.org/abs/hep-th/0111157} {arXiv:hep-th/0111157} \BibitemShut
  {NoStop}%
\bibitem [{\citenamefont {Roy}\ and\ \citenamefont {Saleur}(2022)}]{Roy2021a}%
  \BibitemOpen
  \bibfield  {author} {\bibinfo {author} {\bibfnamefont {A.}~\bibnamefont
  {Roy}}\ and\ \bibinfo {author} {\bibfnamefont {H.}~\bibnamefont {Saleur}},\
  }\bibfield  {title} {\bibinfo {title} {{Entanglement Entropy in the Ising
  Model with Topological Defects}},\ }\href
  {https://doi.org/10.1103/PhysRevLett.128.090603} {\bibfield  {journal}
  {\bibinfo  {journal} {Phys. Rev. Lett.}\ }\textbf {\bibinfo {volume} {128}},\
  \bibinfo {pages} {090603} (\bibinfo {year} {2022})},\ \Eprint
  {https://arxiv.org/abs/2111.04534} {arXiv:2111.04534 [hep-th]} \BibitemShut
  {NoStop}%
\bibitem [{\citenamefont {Rogerson}\ \emph {et~al.}(2022)\citenamefont
  {Rogerson}, \citenamefont {Pollmann},\ and\ \citenamefont
  {Roy}}]{Rogerson:2022yim}%
  \BibitemOpen
  \bibfield  {author} {\bibinfo {author} {\bibfnamefont {D.}~\bibnamefont
  {Rogerson}}, \bibinfo {author} {\bibfnamefont {F.}~\bibnamefont {Pollmann}},\
  and\ \bibinfo {author} {\bibfnamefont {A.}~\bibnamefont {Roy}},\ }\bibfield
  {title} {\bibinfo {title} {{Entanglement entropy and negativity in the Ising
  model with defects}},\ }\href {https://doi.org/10.1007/JHEP06(2022)165}
  {\bibfield  {journal} {\bibinfo  {journal} {JHEP}\ }\textbf {\bibinfo
  {volume} {06}},\ \bibinfo {pages} {165}},\ \Eprint
  {https://arxiv.org/abs/2204.03601} {arXiv:2204.03601 [hep-th]} \BibitemShut
  {NoStop}%
\bibitem [{\citenamefont {Buican}\ and\ \citenamefont
  {Gromov}(2017)}]{Buican2017}%
  \BibitemOpen
  \bibfield  {author} {\bibinfo {author} {\bibfnamefont {M.}~\bibnamefont
  {Buican}}\ and\ \bibinfo {author} {\bibfnamefont {A.}~\bibnamefont
  {Gromov}},\ }\bibfield  {title} {\bibinfo {title} {{Anyonic Chains,
  Topological Defects, and Conformal Field Theory}},\ }\href
  {https://doi.org/10.1007/s00220-017-2995-6} {\bibfield  {journal} {\bibinfo
  {journal} {Commun. Math. Phys.}\ }\textbf {\bibinfo {volume} {356}},\
  \bibinfo {pages} {1017} (\bibinfo {year} {2017})},\ \Eprint
  {https://arxiv.org/abs/1701.02800} {arXiv:1701.02800 [hep-th]} \BibitemShut
  {NoStop}%
\bibitem [{\citenamefont {Houzet}\ and\ \citenamefont
  {Glazman}(2019)}]{Houzet2019}%
  \BibitemOpen
  \bibfield  {author} {\bibinfo {author} {\bibfnamefont {M.}~\bibnamefont
  {Houzet}}\ and\ \bibinfo {author} {\bibfnamefont {L.~I.}\ \bibnamefont
  {Glazman}},\ }\bibfield  {title} {\bibinfo {title} {Microwave spectroscopy of
  a weakly pinned charge density wave in a superinductor},\ }\href
  {https://doi.org/10.1103/PhysRevLett.122.237701} {\bibfield  {journal}
  {\bibinfo  {journal} {Phys. Rev. Lett.}\ }\textbf {\bibinfo {volume} {122}},\
  \bibinfo {pages} {237701} (\bibinfo {year} {2019})}\BibitemShut {NoStop}%
\bibitem [{\citenamefont {Yamamoto}\ \emph {et~al.}(2021)\citenamefont
  {Yamamoto}, \citenamefont {Glazman},\ and\ \citenamefont
  {Houzet}}]{Houzet2021}%
  \BibitemOpen
  \bibfield  {author} {\bibinfo {author} {\bibfnamefont {T.}~\bibnamefont
  {Yamamoto}}, \bibinfo {author} {\bibfnamefont {L.~I.}\ \bibnamefont
  {Glazman}},\ and\ \bibinfo {author} {\bibfnamefont {M.}~\bibnamefont
  {Houzet}},\ }\bibfield  {title} {\bibinfo {title} {Transmission of waves
  through a pinned elastic medium},\ }\href
  {https://doi.org/10.1103/PhysRevB.103.224211} {\bibfield  {journal} {\bibinfo
   {journal} {Phys. Rev. B}\ }\textbf {\bibinfo {volume} {103}},\ \bibinfo
  {pages} {224211} (\bibinfo {year} {2021})}\BibitemShut {NoStop}%
\bibitem [{\citenamefont {Bell}\ \emph {et~al.}(2018)\citenamefont {Bell},
  \citenamefont {Douçot}, \citenamefont {Gershenson}, \citenamefont {Ioffe},\
  and\ \citenamefont {Petković}}]{Bell2018}%
  \BibitemOpen
  \bibfield  {author} {\bibinfo {author} {\bibfnamefont {M.~T.}\ \bibnamefont
  {Bell}}, \bibinfo {author} {\bibfnamefont {B.}~\bibnamefont {Douçot}},
  \bibinfo {author} {\bibfnamefont {M.~E.}\ \bibnamefont {Gershenson}},
  \bibinfo {author} {\bibfnamefont {L.~B.}\ \bibnamefont {Ioffe}},\ and\
  \bibinfo {author} {\bibfnamefont {A.}~\bibnamefont {Petković}},\ }\bibfield
  {title} {\bibinfo {title} {Josephson ladders as a model system for 1d quantum
  phase transitions},\ }\href
  {https://doi.org/https://doi.org/10.1016/j.crhy.2018.09.002} {\bibfield
  {journal} {\bibinfo  {journal} {Comptes Rendus Physique}\ }\textbf {\bibinfo
  {volume} {19}},\ \bibinfo {pages} {484} (\bibinfo {year} {2018})},\ \bibinfo
  {note} {quantum simulation / Simulation quantique}\BibitemShut {NoStop}%
\bibitem [{\citenamefont {Hauschild}\ and\ \citenamefont
  {Pollmann}(2018)}]{Hauschild2018}%
  \BibitemOpen
  \bibfield  {author} {\bibinfo {author} {\bibfnamefont {J.}~\bibnamefont
  {Hauschild}}\ and\ \bibinfo {author} {\bibfnamefont {F.}~\bibnamefont
  {Pollmann}},\ }\bibfield  {title} {\bibinfo {title} {{Efficient numerical
  simulations with Tensor Networks: Tensor Network Python (TeNPy)}},\ }\href
  {https://doi.org/10.21468/SciPostPhysLectNotes.5} {\bibfield  {journal}
  {\bibinfo  {journal} {SciPost Phys. Lect. Notes}\ ,\ \bibinfo {pages} {5}}
  (\bibinfo {year} {2018})}\BibitemShut {NoStop}%
\end{thebibliography}%
\end{document}